\documentclass[letterpaper,english,aps,pra,floatfix,twocolumn]{revtex4}
\usepackage{amsmath}
\usepackage{graphicx}

\makeatletter
\@ifundefined{textcolor}{}
{%
 \definecolor{BLACK}{gray}{0}
 \definecolor{WHITE}{gray}{1}
 \definecolor{RED}{rgb}{1,0,0}
 \definecolor{GREEN}{rgb}{0,1,0}
 \definecolor{BLUE}{rgb}{0,0,1}
 \definecolor{CYAN}{cmyk}{1,0,0,0}
 \definecolor{MAGENTA}{cmyk}{0,1,0,0}
 \definecolor{YELLOW}{cmyk}{0,0,1,0}
 }

\makeatother

\usepackage{babel}

\begin{document}
\global\long\def\avg#1{\langle#1\rangle}

\global\long\def\p{\prime}

\global\long\def\dg{\dagger}

\global\long\def\ket#1{|#1\rangle}

\global\long\def\bra#1{\langle#1|}

\global\long\def\proj#1#2{|#1\rangle\langle#2|}

\global\long\def\inner#1#2{\langle#1|#2\rangle}

\global\long\def\tr{\mathrm{tr}}

\global\long\def\pd#1#2{\frac{\partial#1}{\partial#2}}

\global\long\def\spd#1#2{\frac{\partial^{2}#1}{\partial#2^{2}}}

\global\long\def\der#1#2{\frac{d#1}{d#2}}

\global\long\def\im{\imath}

\global\long\def\As{{^{\sharp}}\hspace{-1mm}\mathcal{A}}

\global\long\def\Fs{{^{\sharp}}\hspace{-0.7mm}\mathcal{F}}

\global\long\def\Es{{^{\sharp}}\hspace{-0.5mm}\mathcal{E}}

\global\long\def\Fd{{^{\sharp}}\hspace{-0.7mm}\mathcal{F}_{\delta}}

\global\long\def\S{\mathcal{S}}

\global\long\def\A{\mathcal{A}}

\global\long\def\F{\mathcal{F}}

\global\long\def\E{\mathcal{E}}

\global\long\def\SgF{\S d\F}

\global\long\def\SgEF{\S d\left(\E/\F\right)}

\global\long\def\U{\mathcal{U}}

\global\long\def\V{\mathcal{V}}

\global\long\def\H{\mathbf{H}_{\S\E}}

\global\long\def\SO{\Pi_{\S}}

\global\long\def\EO{\Upsilon_{k}}

\global\long\def\HSF{\mathbf{H}_{\S\F}}

\global\long\def\HEF{\mathbf{H}_{\S\E/\F}}

\global\long\def\ES{H_{\S}(t)}

\global\long\def\ESo{H_{\S}(0)}

\global\long\def\EgF{H_{\SgF} (t)}

\global\long\def\EgE{H_{\S d\E}(t)}

\global\long\def\EgEF{H_{\SgEF} (t)}

\global\long\def\EF{H_{\F}(t)}

\global\long\def\EFo{H_{\F}(0)}

\global\long\def\ESF{H_{\S\F}(t)}

\global\long\def\ESEF{H_{\S\E/\F}(t)}

\global\long\def\ESSEF{H_{\tilde{\S}\S\E/\F}(t)}

\global\long\def\EEFo{H_{\E/\F}(0)}

\global\long\def\EEF{H_{\E/\F}(t)}

\global\long\def\MI{I\left(\S:\F\right)}

\global\long\def\BS{\left\{  \Pi_{\S}\right\}  }

\global\long\def\QD{\boldsymbol{\delta}\left(\S:\F\right)_{\BS}}

\global\long\def\JI{J\left(\S:\F\right)_{\BS}}

\global\long\def\QDz{\boldsymbol{\delta}\left(\S:\F\right)_{\left\{  \sigma_{\S}^{z}\right\}  }}

\global\long\def\NQD{\bar{\boldsymbol{\delta}}\left(\S:\F\right)_{\BS}}

\global\long\def\EFS{H_{\F\left| \BS\right. }(t)}

\global\long\def\rhoS{\rho_{\S}(t)}

\global\long\def\rhoSo{\rho_{\S}(0)}

\global\long\def\rhoSF{\rho_{\S\F} (t)}

\global\long\def\rhoSgEF{\rho_{\SgEF} (t)}

\global\long\def\rhoSgF{\rho_{\SgF} (t)}

\global\long\def\rhoF{\rho_{\F}(t)}

\global\long\def\rhoFp{\rho_{\F}(\pi/2)}

\global\long\def\LE{\Lambda_{\E}(t)}

\global\long\def\LEc{\Lambda_{\E}^{\star}(t)}

\global\long\def\LF{\Lambda_{\F}(t)}

\global\long\def\LFc{\Lambda_{\F}^{\star}(t)}

\global\long\def\LEF{\Lambda_{\E/\F} (t)}

\global\long\def\LEFc{\Lambda_{\E/\F}^{\star}(t)}

\global\long\def\Hb{H}

\global\long\def\kE{\kappa_{\E}(t)}

\global\long\def\kEF{\kappa_{\E/\F}(t)}

\global\long\def\kF{\kappa_{\F}(t)}

\global\long\def\ts{t=\pi/2}

\global\long\def\mc#1{\mathcal{#1}}

\renewcommand{\onlinecite}[1]{\cite{#1}}

\title{Quantum Darwinism in non-ideal environments}

\author{Michael Zwolak, H. T. Quan, Wojciech H. Zurek}

\affiliation{Theoretical Division, MS-B213, Los Alamos National Laboratory, Los
Alamos, NM 87545}
\begin{abstract}
Quantum Darwinism provides an information-theoretic framework for
the emergence of the objective, classical world from the quantum substrate.
The key to this emergence is the proliferation of redundant information
throughout the environment where observers can then intercept it.
We study this process for a purely decohering interaction when the
environment, $\E$, is in a non-ideal (e.g., mixed) initial state.
In the case of good decoherence, that is, after the pointer states
have been unambiguously selected, the mutual information between the
system, $\S$, and an environment fragment, $\F$, is given solely
by $\F$'s entropy increase. This demonstrates that the environment's
capacity for recording the state of $\S$ is directly related to its
ability to increase its entropy. Environments that remain nearly invariant
under the interaction with $\S$, either because they have a large
initial entropy or a misaligned initial state, therefore have a diminished
ability to acquire information. To elucidate the concept of good decoherence,
we show that - when decoherence is not complete - the deviation of
the mutual information from $\F$'s entropy change is quantified by
the quantum discord, i.e., the excess mutual information between $\S$
and $\F$ is information regarding the initial coherence between pointer
states of $\S$. In addition to illustrating these results with a
single qubit system interacting with a multi-qubit environment, we
find scaling relations for the redundancy of information acquired
by the environment that display a universal behavior independent of
the initial state of $\S$. Our results demonstrate that Quantum Darwinism
is robust with respect to non-ideal initial states of the environment:
the environment almost always acquires redundant information about
the system but its rate of acquisition can be reduced. 
\end{abstract}
\maketitle

\section{Introduction}

Quantum mechanics was initially devised as a microscopic theory of
atoms. However, macroscopic objects are made of quantum components.
Thus quantum mechanics should describe our classical world as well.
Yet, we do not observe ``strange'' quantum states in objects directly
accessible to our senses. This has been a concern since the inception
of quantum mechanics, even as its predictions continue to be verified.
For many years, the strategy was - following Bohr - to bypass this
difficulty by postulating a division between the classical and quantum
worlds \cite{Bohr28-1,Bohr35-1,Wheeler83-1,Zurek91-1}. 

The theory of decoherence is now the standard starting point for addressing
these questions \cite{Zurek91-1,Zurek03-1,Joos03-1,Schlosshauer08-1}.
A system coupled to an environment gets decohered into its pointer
states \cite{Zurek81-1,Zurek03-1} that survive the interaction with
the environment. This durability is one aspect of classicality. Amplification
was also conjectured to play a role \cite{Bohr58-1,Zurek82-1}.
Only very recently, however, has this role been made precise by the
concept of redundancy in \emph{Quantum Darwinism}, an information-theoretic
framework for understanding the quantum-classical transition \cite{Ollivier04-1,Blume05-1,Zurek07-1,Brunner08-1,Bennett08-1,Paz09-1,Zwolak09-1}
(see Ref. \cite{Zurek09-1} for a review). Within this framework,
the objective, classical reality of the pointer states arises from
the redundant dissemination of information about them throughout the
environment. Many observers can then independently determine and reach
consensus about the state of the system by intercepting separate fragments
of the environment. This explains the ``objective reality'' of pointer
states. They are not perturbed by measurements on the environment
and, thus, as classical states should, they are immune to our ``finding
out'' what they are. This process of discovery is especially easy
when fragments of the environment do not interact with each other,
e.g., such as photons. To make the analogy to Darwinism: certain states
- the pointer states - ``survive'' the interaction with the environment
and ``procreate'' by imprinting copies of themselves on the environment.

Quantum Darwinism is an extension of the decoherence paradigm, where
now not only is the system of interest, but so is the environment.
It acts as a witness to the state of the system and as a communication
channel, transmitting information to observers. Previous studies on
Quantum Darwinism focused on models where the system and environment
are initially pure \cite{Ollivier04-1,Blume05-1,Paz09-1}.
It is essential, however, to understand how different initial states
influence the ability of the environment to effectively communicate
information. A recent study has begun to examine the effect of starting
with a ``hazy'' environment, i.e., one with some initial entropy.
It was found that fairly hazy environments behave as noisy communication
channels \cite{Zwolak09-1}. Here we go further by examining more
generally how the environment's capacity to transmit information is
determined by its initial state and also distinguish between the transmission
of quantum and classical information about the system (see also a
recent work by Paz and Roncaglia that examines quantum and classical
information in quantum Brownian motion \cite{Paz09-1}).

We first outline, in Sec. \ref{sec:InfoRedund}, the basic concepts
behind Quantum Darwinism, including the mutual information that is
used to compute the redundancy of information about the system in
the environment. In Sec. \ref{sec:InfoCapacity}, we prove, in the
typical case of good decoherence, that the mutual information between
the system and a fragment of the environment is given by the fragment's
entropy increase when the system interacts independently with many
components of the environment. In Sec. \ref{sec:Discord}, we elucidate
the concept of good decoherence by showing that - when decoherence
is not complete - the deviation of the mutual information from the
fragment's entropy change is quantified by the quantum discord \cite{Zurek00-1,Henderson01-1,Ollivier02-1}.
The excess mutual information between the system and the environment
fragment is information about the initial coherent superposition of
pointer states of the system. 

After these general results, in Sec. \ref{sec:Example}, we introduce
a symmetric environment model composed of qubits that we use to illustrate
the analytic results of Sec. \ref{sec:InfoCapacity} and \ref{sec:Discord}.
We demonstrate how classical information proliferates into the environment.
Also, we investigate the dependence of the redundancy of classical
information storage to \emph{hazy} (i.e., mixed) and \emph{misaligned}
(e.g., close to an eigenstate of the interaction Hamiltonian) initial
environment states. Starting with these non-ideal initial conditions
diminishes the environment's capacity to acquire and transmit information.
For example, in a fairly hazy environment, the redundancy behaves
as $1-h$ as $h\to1$, where the haziness, $h$, is the initial entropy
of an environment qubit. That is, it behaves as a noisy communication
channel. For both hazy and misaligned environments we develop scaling
relations for the behavior of the redundancy. These relations show
a universal behavior of the redundancy that is independent of the
initial state of the system. In Appendix \ref{sec:AppQub}, we solve
for the mutual information and discord for several parameter regimes
of the symmetric environment model. In Appendix \ref{sec:RotTech},
we outline a numerically exact procedure for computing the entropies
(that show up in the mutual information) for the model. In Appendix
\ref{sec:Gaussian}, we derive an approximate expression for the mutual
information that elucidates the behavior of the redundancy.

\section{Information and Redundancy\label{sec:InfoRedund}}

Quantum Darwinism recognizes and investigates the ability of the environment
to redundantly record information about a ``system of interest.''
As before \cite{Ollivier04-1,Blume05-1,Zwolak09-1,Paz09-1},
we focus on the mutual information \begin{equation}
I(\S:\F)=\ES+\EF-\ESF\label{eq:MutInfo}\end{equation}
between the system, $\S$, and a fragment $\F$ of the environment
$\E$. Above, $H_{\S}(t)$ and $H_{\F}(t)$ are the von Neumann entropies
at time $t$ of $\S$ and $\F$, respectively, and $H_{\S\F}(t)$
is the joint entropy $\S$ and $\F$. The mutual information between
$\S$ and $\F$ quantifies the correlations between the two. When
$\S$ and $\F$ are initially uncorrelated, $\MI$ gives the total
information $\F$ gained about the state of $\S$. 

We want to investigate how much information $\F$ acquires about $\S$
when they interact and how redundant this information is. We do not
insist on acquiring all of the missing classical information, $H_{\S}$,
about the system: The information deficit $\delta$ is the fraction
of $H_{\S}$ we are prepared to forgo. For a given $\delta$, the
redundancy of information about $\S$ is the maximum number of disjoint
fragments $R_{\delta}$ that have a mutual information greater than
$\left(1-\delta\right)H_{\S}$ with $\S$. In terms of a fragment
size, the redundancy is \begin{equation}
R_{\delta}=\frac{\Es}{\Fd}=\frac{1}{f_{\delta}},\label{eq:Redund}\end{equation}
where the environment has $\Es$ components, $\Fd$ is the typical
size of an environment fragment needed to acquire a mutual information
no less than $\left(1-\delta\right)H_{\S}$, and $f_{\delta}=\Es/\Fd$
is the corresponding fraction of the environment. In the symmetric
environment considered in Sec. \ref{sec:Example} all possible partitions
of the environment into fragments of size $\Fd$ have identical mutual
information and, thus, $\Fd$ is the size of the environment fragment
needed to give $\MI\ge\left(1-\delta\right)H_{\S}$.

\begin{figure}
\begin{centering}
\includegraphics[width=8cm]{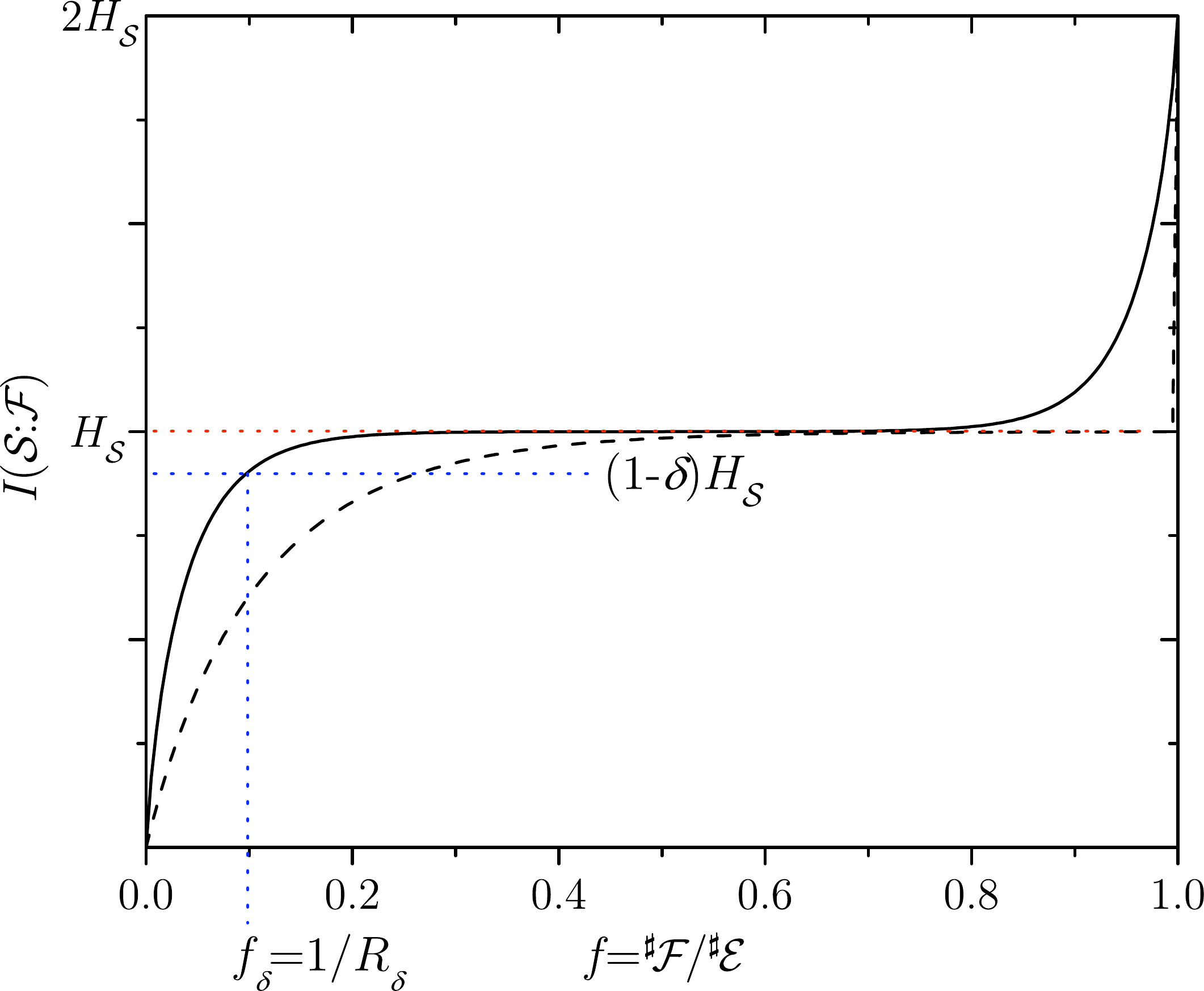}
\par\end{centering}

\caption{The behavior of the mutual information, $\MI$, between the system,
$\S$, and a fragment, $\F$, of a symmetric environment, $\E$, as
a function of the fraction of the environment intercepted, $f=\Fs/\Es$.
The black solid line is for an initially pure $\E$ and the black
dashed line is for an initially hazy $\E$. Here, $\Fs$ and $\Es$
are the number of components in the fragment and the environment,
respectively. The information $H_{\S}$ sets the limit of classical
information about $\S$ that $\E$ can acquire, resulting in a plateau
region at $H_{\S}$ (the \emph{classical plateau}) that signifies
the redundant proliferation of classical information into the environment.
We define a fragment size, $\Fd$, that gives the value of the mutual
information $\left(1-\delta\right)H_{\S}$, i.e., to within the information
deficit, $\delta$, of the classical plateau. The redundancy, $R_{\delta}$,
at this information deficit is then given by Eq. \eqref{eq:Redund}.
The initial haziness of $\E$ reduces the redundancy. The data for
this figure is from an actual simulation like those performed in Sec.
\ref{sec:Example}.\label{fig:Illu}}

\end{figure}

The mutual information given by Eq. \eqref{eq:MutInfo} and redundancy
given by Eq. \eqref{eq:Redund} set the stage for studying how information
is acquired by the environment. Previous studies have shown the formation
of a \emph{classical plateau} with $\MI\simeq H_{\S}$.\emph{ }Figure
\ref{fig:Illu} shows an example of this type of behavior and clarifies
the quantities involved in defining the redundancy.

\section{Information capacity of a purely decohering environment\label{sec:InfoCapacity}}

In this section and the following section, we prove two general results
about how purely decohering environments store and transmit information.
We consider a general model of pure decoherence given by the Hamiltonian
\begin{equation}
\H=\sum_{k=1}^{\Es}\SO\EO,\label{eq:HSE}\end{equation}
where $\SO$ is a Hermitian operator on $\S$ and $\EO$ is a Hermitian
operator on the $k^{th}$ environment component %
\footnote{We assume that $\SO$ and the $\EO$ do not have any degenerate eigenvalues.%
}. This Hamiltonian does not generate transitions between the pointers
states, given by the eigenstates of $\SO$, of the system. In this
model of many environment components interacting independently with
$\S$, we consider a product initial state \begin{equation}
\rho\left(0\right)=\rho_{\S}\left(0\right)\otimes\left[\bigotimes_{k=1}^{\Es}\rho_{k}\left(0\right)\right].\label{eq:InitState}\end{equation}
To compute the mutual information, we calculate the entropies of $\rho_{\S}\left(t\right)$,
$\rho_{\F}\left(t\right)$, and $\rho_{\S\F}\left(t\right)$. Here,
however, we want to first show how the mutual information can be written
more transparently by replacing the entropy of $\rho_{\S\F}(t)$ with
the sum of two entropies: $\rho_{\S}$ decohered only by the remainder
of the environment, $\E/\F$, and $\rho_{\F}\left(0\right)$. That
is, the state \begin{widetext} \begin{eqnarray}
\rho_{\S\F}\left(t\right) & = & \tr_{\E/\F}\left[e^{-\im\H t}\rho\left(0\right)e^{\im\H t}\right]\nonumber \\
 & = & e^{-\im\HSF t}\left\{ \tr_{\E/\F}\left[e^{-\im\HEF t}\rho_{\S\E/\F}\left(0\right)e^{\im\HEF t}\right]\otimes\rho_{\F}\left(0\right)\right\} e^{\im\HSF t},\label{eq:GDproof}\end{eqnarray}
\end{widetext}where $\HSF=\sum_{k\in\F}\SO\EO$ and $\HEF=\sum_{k\in\E/\F}\SO\EO$,
has the same entropy as \begin{equation}
\rhoSgEF\otimes\rho_{\F}\left(0\right).\end{equation}
Here $\SgEF$ is the system decohered solely by $\E/\F$ (i.e., evolved
only by the Hamiltonian $\HEF$) and $\rho_{\F}\left(0\right)=\bigotimes_{k\in\F}\rho_{k}\left(0\right)$
is the initial state of $\F$. Hence the entropy of $\rho_{\S\F}(t)$
is \begin{equation}
H_{\S\F}(t)=\EgEF+\EFo.\label{eq:ESF}\end{equation}
Therefore, the mutual information is \begin{equation}
\MI=\left[\EF-\EFo\right]+\left[\EgE-\EgEF\right],\label{eq:NewMI}\end{equation}
where $\EgE=\ES$, i.e., $\EgE$ is $\S$ decohered by the whole environment
$\E$ %
\footnote{For a pure initial system and environment, $\MI=\EgF+\left[\EgE-\EgEF\right]$,
where the entropy of $\F$ is equivalent to the entropy of $\S$ when
it only interacts with $\F$. There is a generalization to mixed initial
$\S$ and pure initial $\E$, which we discuss later in the paper. %
}. The first term in brackets in Eq. \eqref{eq:NewMI} is the entropy
increase of the fragment $\F$ due to the interaction with $\S$.
The second term is the difference of the entropy of $\S$ interacting
with all of $\E$ and the entropy of $\S$ interacting solely with
$\E/\F$. When both $\E$ and $\E/\F$ are sufficient to decohere
$\S$ at a given time, the second term, $\EgE-\EgEF$, will be nearly
zero. This will happen when $\E$ has decohered $\S$ and the size
of $\F$ is small compared to the size of $\E$. This approximation
of \emph{good decoherence} is accurate at all but very short times
(i.e., less than the decoherence time ) or for very large fragments
(i.e., when the size of $\E/\F$ is too small to decohere $\S$).
Thus, in the typical case of good decoherence, the mutual information
will be approximately \begin{equation}
\MI\approx\EF-\EFo.\label{eq:GDMI}\end{equation}
This reduces to just $\MI\approx H_{\F}\left(t\right)$ for initially
pure environments \cite{Zurek07-1}. The mutual information rewritten
as in Eq. \eqref{eq:GDMI} is a universal relationship for any ``decoherence
only'' model where $\S$ interacts with independent environment components
and where good decoherence has taken place. 

From Eq. \eqref{eq:GDMI}, it is clear that when $\F$ starts in a
state that commutes with $\HSF$, i.e., diagonal in the basis of the
interaction operator that appears in $\HSF$ (either because it is
mixed in that basis or starts in one of the eigenstates of that basis),
it has no capacity to increase its entropy and therefore no capacity
to store classical information about $\S$. In other words, states
of $\E$ that remain invariant under the Hamiltonian dynamics generated
by Eq. \eqref{eq:HSE} do not redundantly store information about
$\S$. The extent to which the environment's initial state coincides
with such states degrades its transmission capabilities.

\section{Discord and Decoherence\label{sec:Discord}}

In this section, we show that before good decoherence has been reached
(or for sufficiently large $\F$), the second term in Eq. \eqref{eq:NewMI}
contributes to the mutual information. This second term is the quantum
discord \cite{Zurek00-1,Henderson01-1,Ollivier02-1} with respect
to the pointer basis of $\S$. The quantum discord with respect to
any basis, $\BS$, is defined as the difference between two classically
equivalent expressions for the mutual information \cite{Ollivier02-1}:\begin{eqnarray}
\QD & = & \MI-\JI\\
 & = & \ES-\ESF+\EFS.\end{eqnarray}
Above, \begin{equation}
\JI=\EF-\EFS\end{equation}
is the other classical expression for the mutual information in terms
of the conditional information (i.e., the entropy decrease of $\F$
given a measurement of $\SO$ on $\S$) %
\footnote{Note that both the information deficit and the discord are denoted
by the same symbol, $\delta$. It should be clear from context to
which quantity $\delta$ refers. However, to help alleviate confusion,
we use a bold $\boldsymbol{\delta}$ for the discord.%
}. 

The second term in brackets in Eq. \eqref{eq:NewMI} is the quantum
discord with respect to the pointer basis of $\S$, i.e., the eigenbasis
of $\SO$ from the Hamiltonian in Eq. \eqref{eq:HSE} %
\footnote{We are not minimizing the discord with respect to the measurement
on $\S$ as we want to differentiate between the information the environment
acquires about the pointer basis and the complementary information
that flows into the environment.%
}. To show this, we first rewrite the quantum discord using Eq. \eqref{eq:ESF}
as \begin{eqnarray}
\QD & = & \EgE-\EgEF\nonumber \\
 &  & -\EFo+\EFS.\end{eqnarray}
The last term, however, simplifies to \begin{eqnarray}
\EFS & = & \sum_{j}p_{j}H_{\F\left|\Pi_{j}^{\S}\right.}(t)\\
 & = & \sum_{j}p_{j}H\left(\U_{j}\rho_{\F}\left(0\right)\U_{j}^{\dagger}\right)\\
 & = & \EFo,\end{eqnarray}
where $p_{j}$ is the occupation of the $j^{th}$ eigenstate of $\SO$
and $\U_{j}$ is the evolution operator projected onto that state.
Thus, in this case of pure decoherence by independent environment
components, the quantum discord is \begin{equation}
\QD=\EgE-\EgEF.\label{eq:QD}\end{equation}
The discord represents information complementary to the information
about the pointer states of $\S$ that the environment fragment has
acquired. To see this, note that the discord in Eq. \eqref{eq:QD}
involves only the entropy of $\S$ evolved in the presence of the
full environment $\E$ and the environment without the fragment $\E/\F$.
Under a pure decoherence Hamiltonian, any difference between these
two is due to off-diagonal elements in the system's initial density
matrix. That is, the discord yields information about the initial
coherence between pointer states of $\S$. This is information about
the complementary observables to $\SO$, i.e., operators that do not
commute with $\SO$. 

In pure decoherence models, the same complementary information flows
into the environment regardless of whether $\E$ is in an initially
pure or mixed state. This comes out of Eq. \eqref{eq:QD} after recognizing
that the environment decoheres the system identically regardless of
its initial entropy when its alignment is held fixed.. However, even
though the initial entropy of the environment does not effect its
ability to receive complementary information, its alignment with the
states that commute with $\H$ does effect this ability. These issues
will be discussed along with the following concrete, solvable example
in order to elucidate the ideas shown here.

\section{Example: Qubit interacting with a symmetric environment\label{sec:Example}}

\begin{figure}
\begin{centering}
\includegraphics[width=8cm]{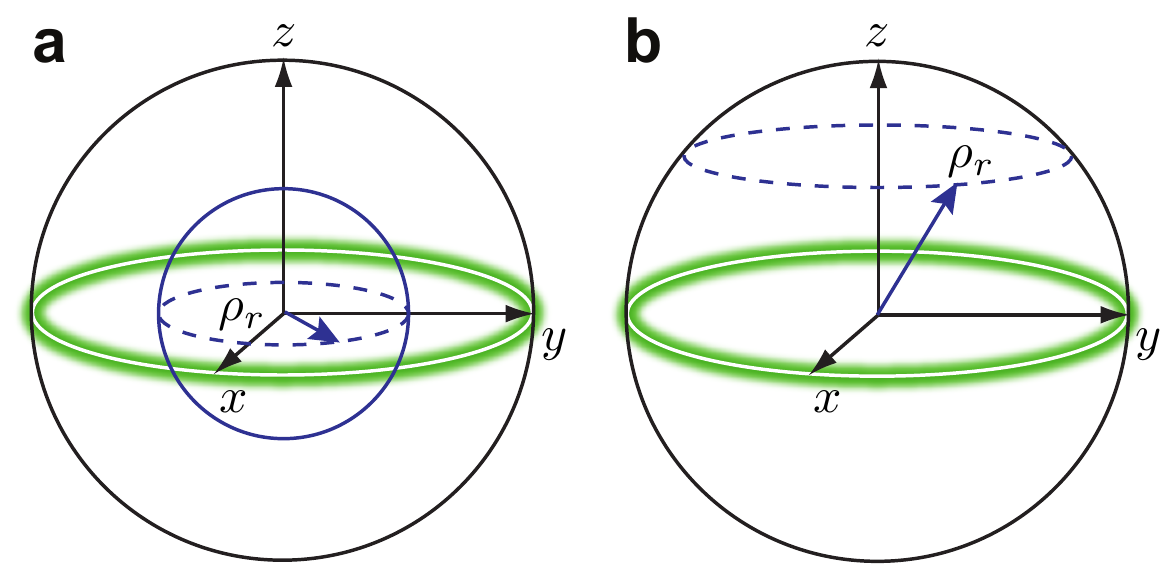}
\par\end{centering}

\caption{Bloch sphere representation of (a) haziness and (b) misalignment.
Pure states in the x-y plane (highlighted) have the maximum capacity
to accept information about the system's pointer states. Haziness
of an environment qubit contracts its Bloch sphere, reducing the qubits
ability to increase its entropy and therefore decreasing its capacity
to store information. Misalignment rotates the state out of the x-y
plane, which also decreases the qubits capacity to store information.
For both haziness and misalignment, the decrease in capacity is due
to a reduction in the environment qubits' ability to branch into two
orthogonal states correlated with the two pointer states of the system.\label{fig:Bloch}}

\end{figure}

We now study a solvable example of a qubit system interacting with
a symmetric qubit environment often used as a model of decoherence
\cite{Zurek81-1,Zurek82-1}. The Hamiltonian is \begin{equation}
\H=\frac{1}{2}\sum_{k=1}^{\Es}\sigma_{\S}^{z}\sigma_{k}^{z}.\label{eq:Hsym}\end{equation}
It causes pure decoherence of the system's state into its pointer
basis - the eigenstates of $\sigma_{\S}^{z}$. In this basis, the
system is initially described by \begin{equation}
\rhoSo=\left(\begin{array}{cc}
s_{00} & s_{01}\\
s_{10} & s_{11}\end{array}\right).\label{eq:rhoSinit}\end{equation}
We take the initial state of the environment to be the product state,
Eq. \eqref{eq:InitState}, with $\rho_{k}\left(0\right)=\rho_{r}$
for all $k$. In the $\sigma^{z}$ basis, the density matrix of each
component is \begin{equation}
\rho_{r}=\left(\begin{array}{cc}
r_{00} & r_{01}\\
r_{10} & r_{11}\end{array}\right).\label{eq:rhor}\end{equation}
We examine how two quantities that characterize this state, its haziness
and misalignment, affect its ability to accept information. Figure
\ref{fig:Bloch} shows a representation of these quantities using
the Bloch sphere. The \emph{haziness} is the preexisting entropy of
an environment qubit:\begin{equation}
h\equiv H\left(\rho_{r}\right).\label{eq:haziness}\end{equation}
\emph{Misalignment} of a component of the environment is tilting it
away from the states that have the most capacity to accept information.
Thus, we can similarly define the misalignment of the environment
by the maximum entropy it can obtain under a pure decoherence Hamiltonian,
\begin{equation}
h_{m}\equiv H\left(r_{00}\right),\label{eq:MisalignHm}\end{equation}
where $H(x)\equiv-x\log_{2}x-\left(1-x\right)\log_{2}(1-x)$ is the
binary entropy. This parameter indicates the maximum amount of information
(according to Eq. \eqref{eq:GDMI}) that an environment qubit can
ever obtain after good decoherence has taken place under the evolution
of Eq. \eqref{eq:Hsym}. The maximum capacity states are qubits that
start in the $x-y$ plane of the Bloch sphere. The minimum capacity
states are $\sigma^{z}$ eigenstates, which will not even decohere
$\S$. When we calculate the redundancy, however, we find it more
convenient to parametrize the misalignment of $\rho_{r}$ as \begin{equation}
\sigma=r_{00}-r_{11}\label{eq:Misparam}\end{equation}
instead of using Eq. \eqref{eq:MisalignHm}. When an environment qubit
is in a $\sigma^{z}$ eigenstate, $\left|\sigma\right|=1$, it will
remain untouched by $\H$ of Eq. \eqref{eq:Hsym}. The details of
the calculations can be found in the appendices. In the following
we highlight the main results.

\subsection{Mutual Information}

\begin{figure*}
\begin{centering}
\includegraphics[width=12cm]{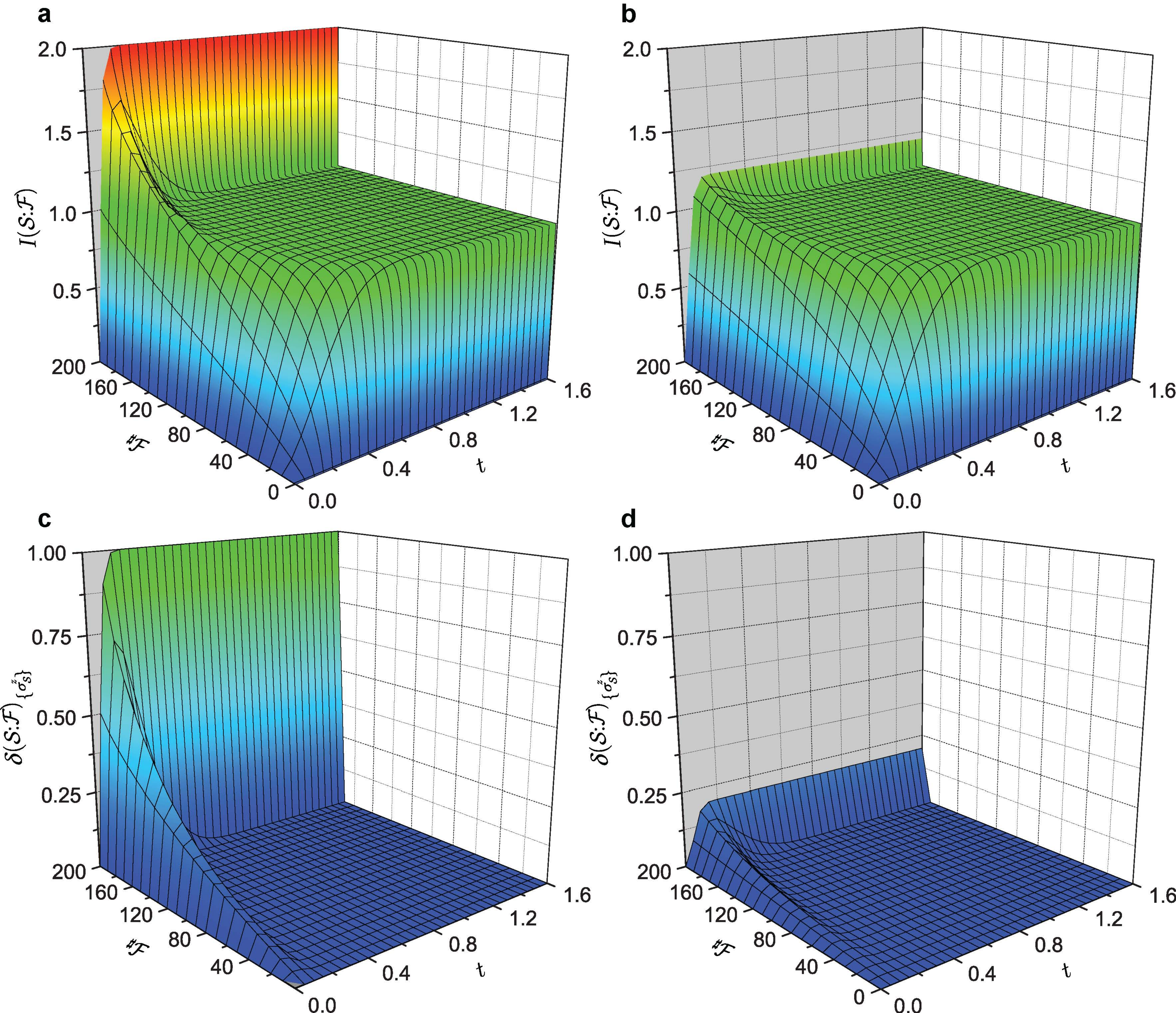}
\par\end{centering}

\caption{(a,b) Mutual information, $\MI$, and (c,d) Quantum discord, $\QDz$,
versus the fragment size, $\Fs$, and time, $t$, with $s_{00}=1/2$,
$r_{00}=1/2$, and $\Es=200$. (a,b) The mutual information for $\ESo=0$
and $\ESo=0.8$, respectively. Initially $\E$ and $\S$ are uncorrelated,
but as time develops, $\E$ acquires information about $\S$. After
an initial transient region - signified by a nonzero quantum discord
in (c,d) - a plateau develops in the mutual information. A sudden
increase in the von Neumann mutual information occurs for large $\Fs\sim\Es$
because complementary information about $\S$ is accessible via global
measurements. For a system initially in a superposition, this jump
is large and so is the discord in the transient region. The jump is
reduced by any existing decoherence of $\S$ when it is placed in
contact with $\E$. However, the level and size of the classical plateau
is identical regardless of the initial entropy of $\S$.\label{fig:MIPureE}
(c,d) The quantum discord with respect to the eigenstates of $\SO=\sigma_{\S}^{z}$
for $\ESo=0$ and $\ESo=0.8$, respectively. The environment, $\E$,
can be pure or mixed, and, in fact, the discord is equivalent to the
mutual information between $\S$ and $\F$ for a diagonal initial
state $\rho_{r}$. There is a transient region, just after $\S$ and
$\E$ have come into contact, where nonzero discord exists. Its duration
depends on the size of the environment. Except for this region, the
discord is negligibly small since both $\E$ and $\E/\F$ are sufficient
to decohere $\S$. The discord is reduced by the preexisting entropy
of $\S$ before coming in contact with $\E$, as shown in (b). This
is because the discord signifies complementary information about $\S$,
i.e., information about the initial coherence between pointer states
of $\S$.\label{fig:QD} }

\end{figure*}

\begin{figure}
\begin{centering}
\includegraphics[width=6cm]{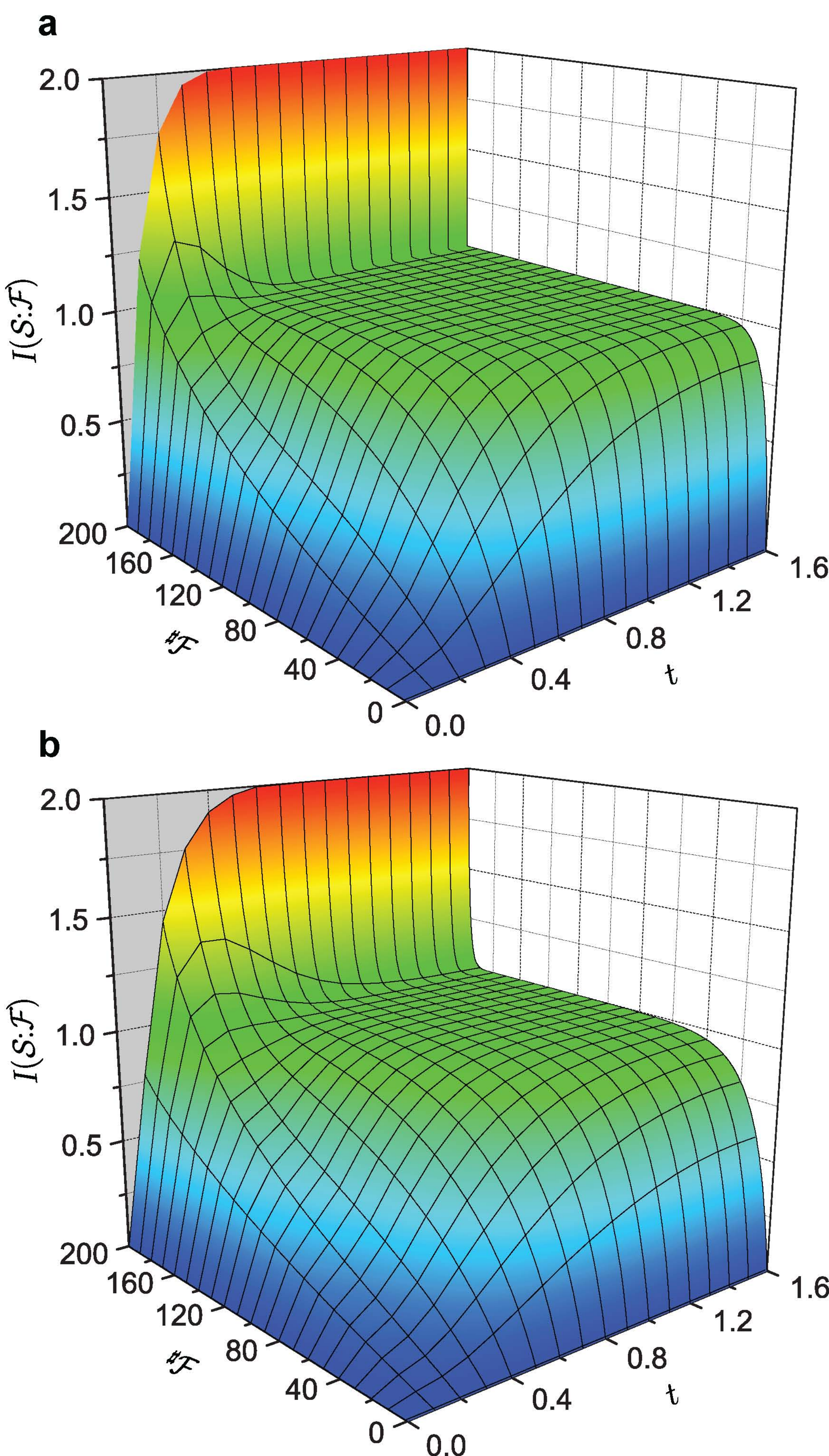}
\par\end{centering}

\caption{Mutual information, $\MI$, versus the fragment size, $\Fs$, and
time, $t$, for an initially pure $\S$ and (a) an initially hazy
$\E$ with $r_{00}=1/2$ and $h\approx0.8$, and (b) an initially
hazy and misaligned $\E$ with $\sigma=0.8$ and $h/h_{m}\approx0.8$.
The other parameters are $s_{00}=1/2$ and $\Es=200$. As with Fig.
\ref{fig:MIPureE}, $\E$ initially contains no information about
$\S$, but correlations develop, transmitting information about $\S$
throughout $\E$, then going over to a classical regime signified
by the formation of the classical plateau. Misalignment and haziness
both reduce the information a given fragment $\F$ gains about $\S$.
However, since the classical plateau forms, the environment still
maintains the ability to redundantly encode classical information
about $\S$.\label{fig:MImixedE}}

\end{figure}

In Appendix \ref{sec:AppQub} and using Eq. \eqref{eq:NewMI}, we
show that the mutual information takes on the form \begin{eqnarray}
I\left(\S:\F\right) & = & \left[H_{\F}\left(t\right)-H_{\F}\left(0\right)\right]\nonumber \\
 &  & +\left[H\left(\kE\right)-H\left(\kEF\right)\right],\label{eq:MIsym}\end{eqnarray}
where \begin{equation}
\kappa_{\mathcal{A}}(t)=\frac{1}{2}\left(1+\sqrt{\left(s_{11}-s_{00}\right)^{2}+4\left|s_{01}\right|^{2}\left|\Lambda_{\mathcal{A}}(t)\right|^{2}}\right)\label{eq:kapE}\end{equation}
and $\Lambda_{\mathcal{A}}(t)$ is the contribution to decoherence
of $\S$ due to the subset $\A$ of the environment. Figures \ref{fig:MIPureE}(a,b)
and \ref{fig:MImixedE}(a,b) show the behavior of the mutual information
versus time for several different cases involving pure and mixed $\S$
and pure, mixed, and misaligned $\E$. 

For pure $\S$ and pure $\E$, the mutual information is plotted in
Fig. \ref{fig:MIPureE}(a). Initially $\S$ and $\E$ are uncorrelated
and therefore the environment contains no information about the system.
In time, however, correlations begin to encode information about both
the pointer states of $\S$ and their superpositions. The latter is
reflected by the nonzero quantum discord in Fig. \ref{fig:QD}(c,d).
After good decoherence has taken place, a plateau develops in the
mutual information as a function of $\Fs$. This \emph{classical plateau}
signifies classical (i.e., redundant and therefore objective) information
that has proliferated throughout the environment. 

For mixed $\S$ and pure $\E$, the mutual information is plotted
in Fig. \ref{fig:MIPureE}(b) for $\ESo=0.8$. As with a pure $\S$,
the environment develops correlations with the system. In particular,
it obtains information about the pointer states of the system. Thus,
as before, the classical plateau forms at the same level, $H_{\S}$,
which is determined only by diagonal elements of the system's initial
density matrix in its pointer basis. However, the available complementary
information about $\S$, as signified by the discord with $\F$, is
reduced due to the initial entropy of $\S$.

Environments, however, will generally contain some preexisting entropy,
e.g., due to a finite temperature or interactions with other degrees
of freedom not directly in contact with $\S$ (for example, photons
emitted from the sun are initially partially mixed). In Fig. \ref{fig:MImixedE},
the mutual information is plotted for a hazy environment, $h\approx0.8$,
and a hazy, misaligned environment with $\sigma=0.8$ and $h/h_{m}\approx0.8$
\footnote{For both the initially hazy $\E$ and the initially hazy, misaligned
$\E$, the initial density matrix $\rho_{r}$ in Eq. \eqref{eq:rhor}
is constructed first by creating a pure state (with $\sigma=0$ and
$\sigma=0.8$, respectively), then creating the initial entropy by
decohering the off-diagonal matrix elements by a factor $1/2$. This
creates an initial $\rho_{r}$ with $h\approx0.8$ and $h/h_{m}\approx0.8$,
respectively, rather than exactly 0.8.%
}. Although the classical plateau is slower to develop as a function
of $\Fs$, it still forms at the same level, $H_{\S}$, as for an
initially pure $\E$. This is significant, as it shows that the pointer
states of $\S$ can still be completely determined from a fragment
of $\E$ even when the environment is initially in a non-ideal state.

\subsection{Discord}

The second term in brackets in Eq. \eqref{eq:MIsym} gives the quantum
discord with respect to the eigenstates of $\sigma_{\S}^{z}$: \begin{equation}
\QDz=H\left(\kE\right)-H\left(\kEF\right),\label{eq:QDsym}\end{equation}
which is plotted in Fig. \ref{fig:QD}(c,d) for two initial conditions.
This is the deviation from the good decoherence expression, Eq. \eqref{eq:GDMI},
for the mutual information. This deviation term will be nearly zero
whenever $\kE\approx\kEF$, which occurs when $\left|\LE-\LEF\right|\approx0$
- that is, whenever both $\E$ and $\E/\F$ are sufficient to decohere
$\S$ %
\footnote{The discord will also be zero when the environment and system are
in a product state, as they are at $t=0$. In this case, Eq. \eqref{eq:GDMI}
holds but only because the system and environment are uncorrelated.%
}. In this symmetric model, good decoherence means that both $\Es$
and $\Es-\Fs$ are sufficiently large, or that $\Lambda_{k}(t)$,
the contribution of a single $\E$ spin to decoherence (see Eq. \eqref{eq:Lamk}),
is sufficiently small, so the decoherence factors $\LE$ and $\LEF$
are both small. 

As discussed above, the discord represents information the environment
fragment has acquired regarding complementary observables of $\S$.
In this qubit system with a $\sigma_{\S}^{z}$ pointer basis, the
complementary observables are $\sigma_{\S}^{x}$ and $\sigma_{\S}^{y}$.
The initial expectation value of these observables are $\avg{\sigma_{\S}^{x}}_{0}=2\Re s_{01}$
and $\avg{\sigma_{\S}^{y}}_{0}=-2\Im s_{01}$, respectively. Since
the discord is the difference of two terms, which only differ by the
factor multiplying $s_{01}$, it contains information regarding the
initial expectation value of $\sigma_{\S}^{x}$ and $\sigma_{\S}^{y}$,
whereas the first term in brackets in Eq. \eqref{eq:MIsym} does not
(as can be seen from the form of $\rhoF$ in Eq. \eqref{eq:rhoF}).
This is more obvious close to good decoherence when the discord becomes
\begin{eqnarray}
\QDz & \approx & \left(\avg{\sigma_{\S}^{x}}_{0}^{2}+\avg{\sigma_{\S}^{y}}_{0}^{2}\right)\times\\
 &  & \left(\left|\LEF\right|^{2}-\left|\LE\right|^{2}\right)\frac{\log_{2}\left(s_{00}/s_{11}\right)}{4\left(s_{11}-s_{00}\right)}.\nonumber \end{eqnarray}
That is, the quantum discord is directly proportional to the expectation
value of the observables that do not commute with the pointer observable
$\sigma_{\S}^{z}$.

Equation \eqref{eq:QDsym} together with Eq. \eqref{eq:kapE} also
show that whether the environment is pure or hazy, it acquires identical
complementary information about $\S$. This is evident by the dependence
of the quantum discord only on how $\E$ and $\E/\F$ decohere $\S$.
The latter only relies on the initial alignment of the environment
components with the eigenstates of $\sigma^{z}$, but not on how hazy
they are.

\subsection{Redundancy}

In the previous two subsections we examined the behavior of the mutual
information and quantum discord in various parameter regimes. We now
examine the behavior of the redundancy for different initial states
of the system and environment.

\begin{figure*}
\begin{centering}
\includegraphics[width=12cm]{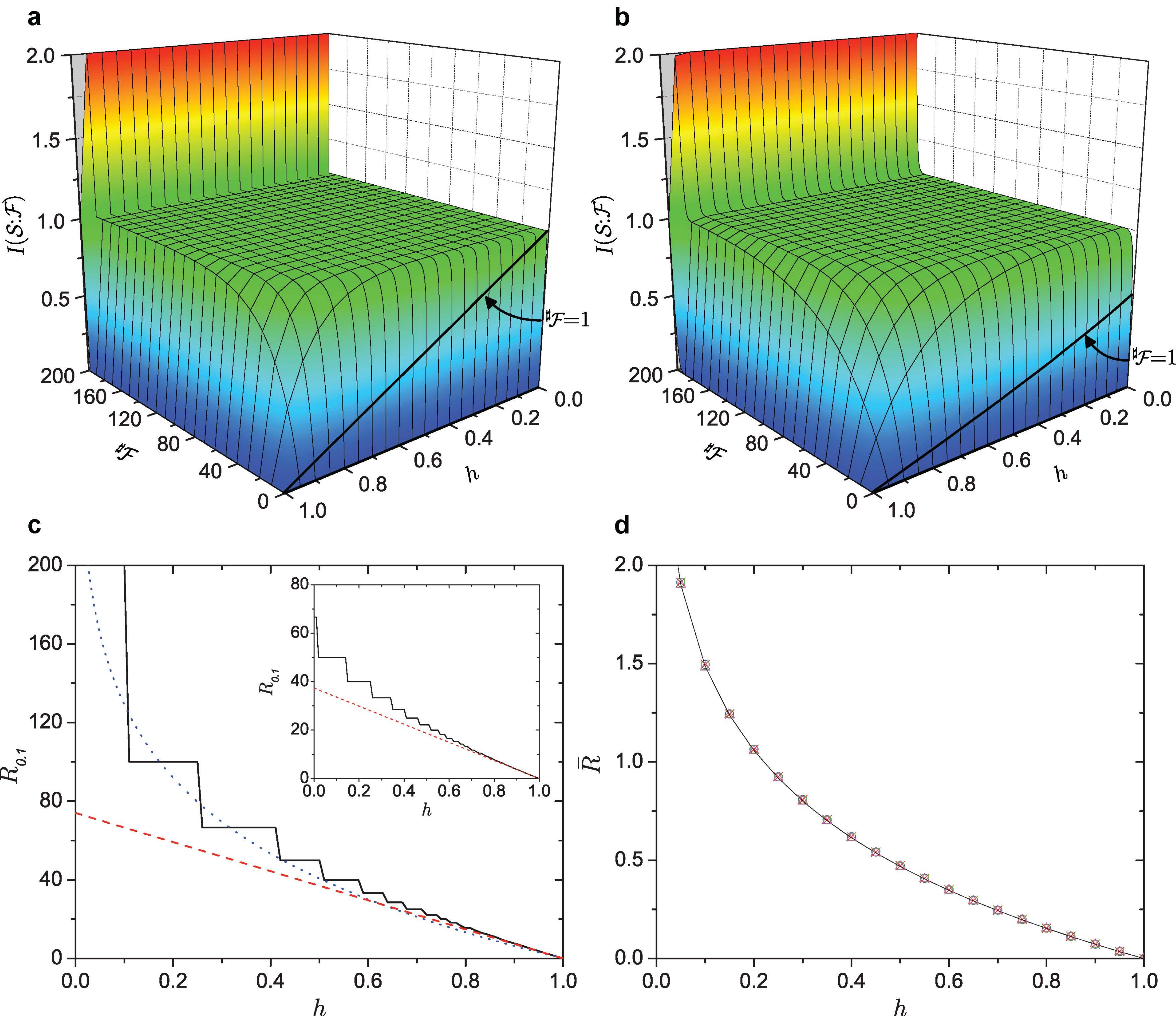}
\par\end{centering}

\caption{(a,b) Mutual information, $\MI$, versus the fragment size, $\Fs$,
and haziness, $h$, of the environment qubits, and (c,d) the redundancy
(limiting redundancy), $R_{\delta}$ ($\bar{R}$), versus $h$. The
system is initially pure, and $s_{00}=1/2$, $r_{00}=1/2$, and $\Es=200$.
(a,b) Mutual information at $\ts$ and $t=\pi/4$, respectively. The
classical plateau forms in all but the haziest of conditions where
the environment is ab initio in a perfect mixture.\label{fig:MImixed}
(c) The redundancy, $R_{\delta}$, at $\ts$ (and $t=\pi/4$ in the
inset). The black line is the exact data. The redundancy initially
drops fairly rapidly because of the symmetry of the model but then
goes over to a region where it behaves as $1-h$ (shown as the red,
dashed line). The blue, dotted curve shows the scaling behavior from
Eq. \eqref{eq:Rscaling}, which already reasonably approximates the
exact behavior even though $\Fd$ is small.\label{fig:Rmixed} (d)
The limiting redundancy, $\bar{R}$, at $t=\pi/2$. There are four
different initial states of $\S$: $s_{00}=1/2$ (blue circles), $s_{00}=1/8$
(red crosses), $s_{00}=1/64$ (green squares), and $s_{00}=1/4096$
(magenta triangles). Equation \eqref{eq:LimitR} with $R_{\delta}$
from Eq. \eqref{eq:Rscaling}, is plotted (black line) with the exact
data for the different initial states of $\S$.\label{fig:RscalingMixed}}

\end{figure*}

\emph{Hazy }$\E$ - As discussed above, an initially hazy $\E$ has
a lower capacity to store information \cite{Zwolak09-1}. In Fig.
\ref{fig:MImixed}(a,b), we plot the mutual information versus $\Fs$
and $h$ at $\ts$ and $t=\pi/4$. Even though the initial haziness
diminishes the capacity of the environment to acquire and transmit
information, we see that the classical plateau still forms and at
the same level ($H_{\S}$), but takes a longer time to develop and
flattens out only for larger $\Fs$. Moreover, the final jump of the
mutual information when $\Fs\approx\Es$, which signifies complete
quantum correlation of $\E$ with $\S$, is the same regardless of
whether the environment is initially pure or hazy. Somewhat surprisingly,
it occurs even for a completely hazy environment ($h=1$) where the
classical plateau is missing. Thus the complementary information about
$\S$ remains the same regardless of the haziness, $h$, at fixed
misalignment.

In Fig. \ref{fig:Rmixed}(c), we plot the redundancy for $\ts$ and
$t=\pi/4$. This shows explicitly that although the capacity of the
environment is reduced, the redundancy is still large. There is an
initial, more rapid drop in the redundancy as the state becomes a
little hazy, but this crosses over to a linear region where redundancy
behaves as $1-h$, i.e., like a noisy communication channel \cite{Cover06-1}.
The initial, more rapid drop at $t=\pi/2$ is due to the symmetry
of the environment: when $h=0$ each qubit has complete classical
correlation with $\S$.

In Appendix \ref{sec:Gaussian}, we derive an approximate expression
for the mutual information at $r_{00}=1/2$ and $\ts$ for fairly
hazy $\E$ and large $\Fs$: \begin{equation}
\MI\approx H_{\S}-\frac{\left(2\sqrt{\lambda_{-}\lambda_{+}}\right)^{\Fs}}{\sqrt{\pi\Fs/2}}\frac{2\pi\sqrt{s_{00}s_{11}}}{\left(\ln2\right)\left(\ln\frac{\lambda_{+}}{\lambda_{-}}\right)}.\label{eq:AppMI}\end{equation}
This asymptotic expression allows us to estimate the redundancy when
the information deficit, $\delta$, is small as \begin{equation}
R_{\delta}\approx\frac{\Es\ln\left(2\sqrt{\lambda_{-}\lambda_{+}}\right)}{\ln\delta}.\label{eq:Rscaling}\end{equation}
This expression is plotted in Fig. \ref{fig:Rmixed}(c) along with
the exact data (and also the linear approximation) for $t=\pi/2$.
Even when $\Fd$ is small (i.e., for small information deficits and
haziness), this approximation captures the behavior of the redundancy
\footnote{We emphasize, however, that the approximation to the mutual information,
Eq. \eqref{eq:AppMI}, from which it is derived, does not work well
at small $\Fs$, as can be seen in Fig. \ref{fig:Errors}.%
}. As $\delta\to0$, the redundancy for an arbitrary initial system
state collapse onto this same universal curve. Thus, we define a limiting
redundancy \begin{equation}
\bar{R}=\lim_{\delta\to0}\frac{-R_{\delta}\ln\delta}{\Es}.\label{eq:LimitR}\end{equation}
This expression, with $R_{\delta}$ from Eq. \eqref{eq:Rscaling},
is shown in Fig. \ref{fig:RscalingMixed}(d) along with $\bar{R}$
from the exact data for four different initial states of $\S$: $s_{00}=1/2$,
$s_{00}=1/8$, $s_{00}=1/64$, and $s_{00}=1/4096$. From the figure,
we see that when discrete effects disappear, the limiting redundancy
describes very well the behavior of the redundancy of information
proliferated into the environment and that this behavior is universal
- it does not depend on the system's initial state. 

\begin{figure*}
\begin{centering}
\includegraphics[width=12cm]{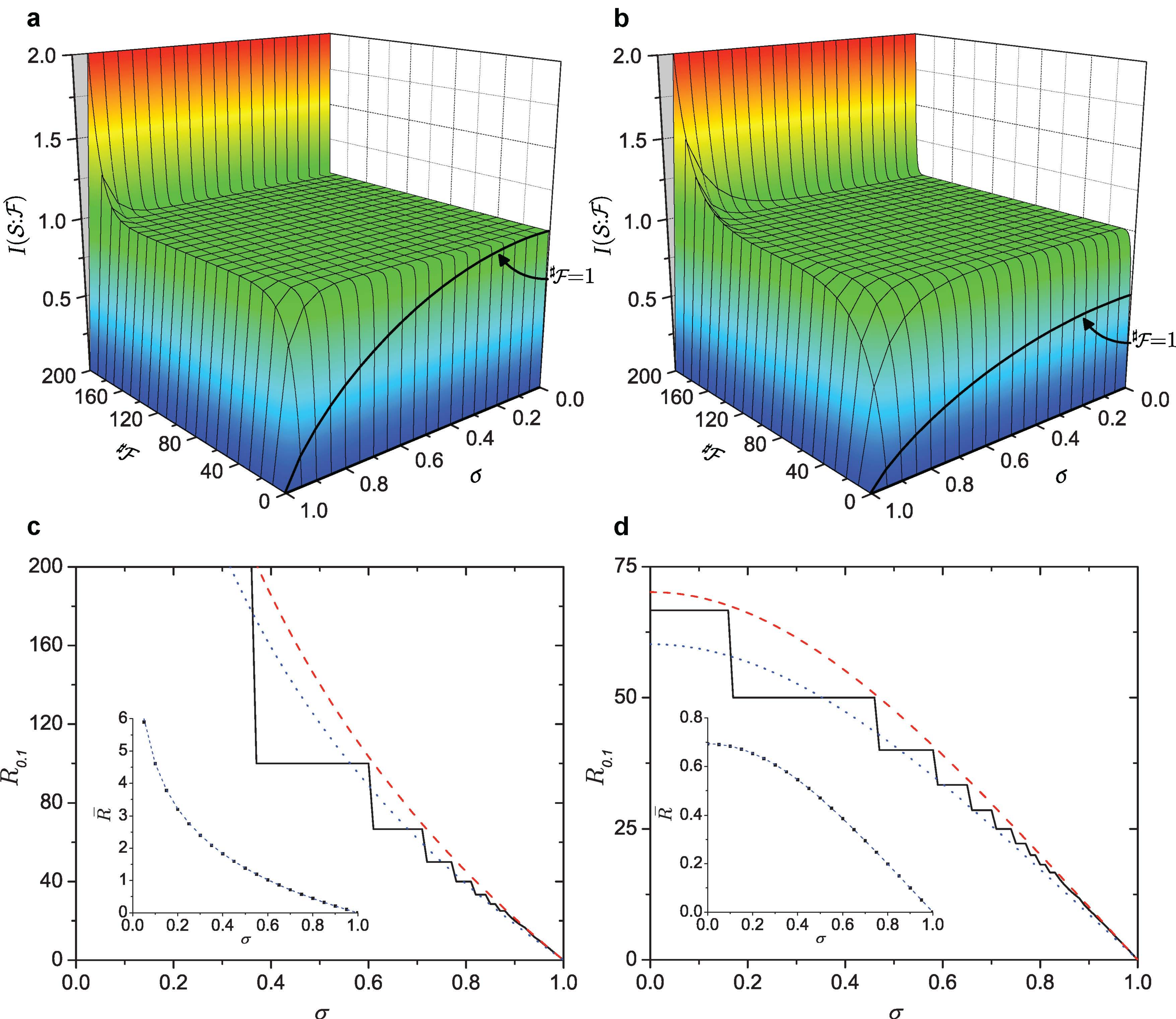}
\par\end{centering}

\caption{(a,b) Mutual information, $\MI$, versus the fragment size, $\Fs$,
and misalignment, $\sigma$, of the environment qubits, and (c,d)
the redundancy, $R_{\delta}$, versus $\sigma$. The system and environment
are initially pure, and $s_{00}=1/2$ and $\Es=200$. (a,b) The mutual
information at $t=\pi/2$ and $t=\pi/4$, respectively. The classical
plateau is formed and is quite large for all but very misaligned states
($\sigma$ near 1). \label{fig:MImis} (c,d) The redundancy versus
the misalignment at $t=\pi/2$ and $t=\pi/4$, respectively. The black
lines are the exact data obtained numerically and the blue dotted
line is the scaling given by Eq. \eqref{eq:RScalingMis}, which already
fits quite well with the numerical results. The red dashed line is
the redundancy given by $R_{\delta}\propto\ln\left|\Lambda_{k}\left(t\right)\right|^{2}$
with the constant of proportionality found by retaining all the factors
when using Eq. \eqref{eq:IappMis}. The inserts are the limiting redundancy,
$\bar{R}$, given by Eq. \eqref{eq:LimitR} (using both the exact
data shown as squares and data from Eq. \eqref{eq:RScalingMis} shown
as a dashed blue line), which demonstrates that the scaling result
is obtained when discrete effects are not present (i.e., in the limit
of vanishing information deficit, $\delta\to0$). \label{fig:Rmis}}

\end{figure*}

\emph{Misaligned} $\E$ - As discussed above, a misaligned environment
qubit is one that has a larger overlap with an eigenstate of the interaction
Hamiltonian, and thus one with a decreased capacity for information.
With the interaction Hamiltonian containing $\sigma^{z}$ operators
on the environment qubits, the misalignment is the bias in the initial
state, Eq. \eqref{eq:rhor}, $\sigma=r_{00}-r_{11}$. In Fig. \ref{fig:MImis}(a,b),
we show the mutual information versus $\Fs$ and $\sigma$ at $\ts$
and $t=\pi/4$. The classical plateau is formed for all but the most
misaligned states and at the same level, $H_{\S}$. Thus, just as
with haziness, misaligned environments also redundantly encode information
(i.e., classical information) about $\S$. The redundancy is plotted
in Fig. \ref{fig:Rmis}(c,d) for these two times. We can see that,
for the not too small information deficit $\delta=0.1$, $R_{\delta}$
is initially quite insensitive to the misalignment.

We can get quantitative understanding of how the redundancy behaves
if we take $\delta$ to be small. In this case, a large $\Fd$ is
necessary to achieve the plateau value of the mutual information within
the information deficit $\delta$ and we thus can take all the corresponding
decoherence factors $\LF$, $\LEF$, and $\LE$ to be very small and
expand the entropies in the mutual information, Eq. \eqref{eq:MIpureE}.
For pure $\E$, as long as $\Es\gg\Fd$, this gives the mutual information\begin{equation}
\MI\approx H_{\S}-\frac{s_{00}s_{11}\ln\frac{s_{00}}{s_{11}}}{\left(s_{00}-s_{11}\right)\ln2}\left|\LF\right|^{2}.\label{eq:IappMis}\end{equation}
Thus, we have \begin{equation}
\Fd\approx\frac{\ln\delta}{\ln\left|\Lambda_{k}\left(t\right)\right|^{2}},\end{equation}
where $\left|\Lambda_{k}\left(t\right)\right|^{2}=\cos^{2}t+\sigma^{2}\sin^{2}t$.
Therefore, the redundancy for small information deficit will scale
as \begin{equation}
R_{\delta}\approx\frac{\Es\ln\left|\Lambda_{k}\left(t\right)\right|^{2}}{\ln\delta}.\label{eq:RScalingMis}\end{equation}
This is plotted in Fig. \ref{fig:Rmis}(c,d) along with the exact
redundancy. Note that, even for the information deficit $\delta=0.1$,
the approximate expression is quite good modulo discrete effects.
The insets in Fig. \ref{fig:Rmis}(c,d) show that as the information
deficit is taken to zero, $\delta\to0$, the scaling predicted for
the limiting redundancy, $\bar{R}$ from Eq. \eqref{eq:LimitR} with
$R_{\delta}$ given by Eq. \eqref{eq:RScalingMis}, describes the
redundancy behavior of misaligned states very well. At $t=\pi/2$,
the redundancy becomes proportional to $\ln\sigma$. Two noticeable
features, which are similar to the scaling for hazy, but aligned,
environments, given by Eq. \eqref{eq:Rscaling}, are that the redundancy
is inversely proportional to the logarithm of the information deficit
and that, for small $\delta$, the redundancy is insensitive to the
alignment (or initial entropy) of the system. This supports the idea
that the redundancy has a universal behavior independent of the system's
initial state.

\begin{figure*}
\begin{centering}
\includegraphics[width=18cm]{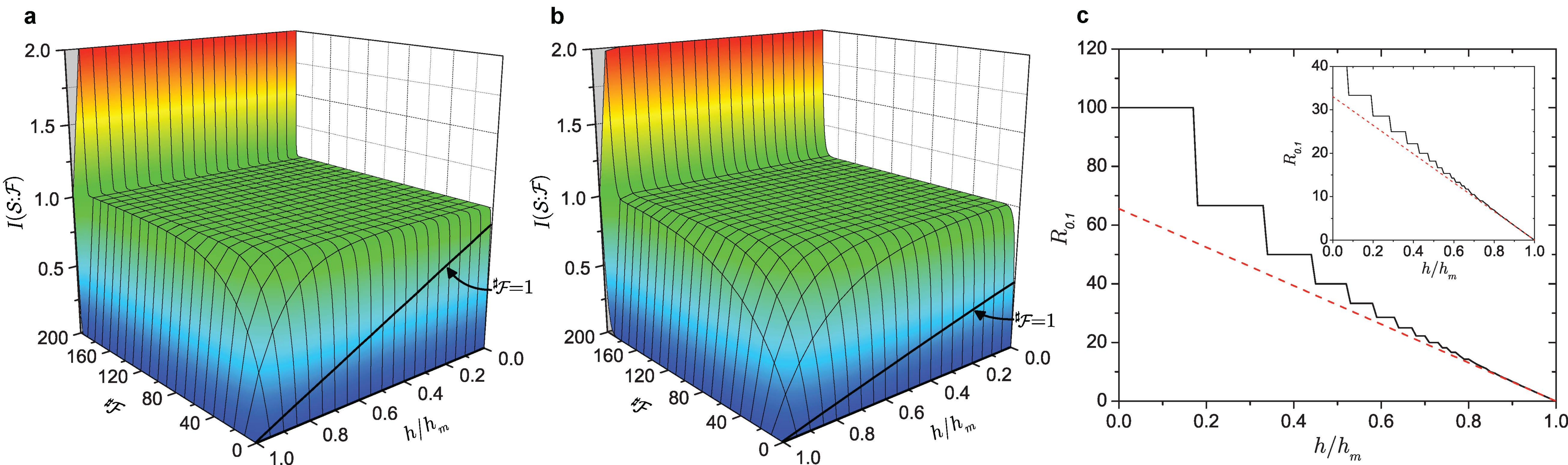}
\par\end{centering}

\caption{(a,b) Mutual information, $\MI$, versus the fragment size, $\Fs$,
and normalized haziness, $h/h_{m}$, of the environment qubits, and
(c) the redundancy, $R_{\delta}$, versus $h/h_{m}$. The system is
initially pure, and $t=\pi/2$, $s_{00}=1/2$, and $\Es=200$. (a,b)
Mutual information for the misalignments $\sigma=0.4$ ($h_{m}\approx0.88$)
and $\sigma=0.8$ ($h_{m}\approx0.47$), respectively. The plateau
region still forms at the same level, $H_{\S}$, essentially regardless
of how hazy or misaligned the environment is initially. \label{fig:MImismixed}
(c) Redundancy for $\sigma=0.4$ (the inset is for $\sigma=0.8$).
The black line is the exact data and the red dashed line is the linear
approximation. As we see, there is still a linear region of the redundancy
for fairly hazy environments. \label{fig:Rmismixed}}

\end{figure*}

\emph{Misaligned and hazy} $\E$ - We now consider the case where
$\E$ is both misaligned and hazy. In Fig. \ref{fig:MImismixed}(a,b),
the mutual information is plotted versus $\Fs$ and $h/h_{m}$ for
$\sigma=0.4$ and $\sigma=0.8$. Just as with misalignment and haziness
separately, one still gets the formation of the classical plateau
and hence, one still gets redundancy. For fairly hazy environments,
the redundancy behaves as for $\sigma=0$ but now with a rescaled
haziness $h/h_{m}$. The quantity $h_{m}$ represents the maximum
information capacity of a single environment qubit given its alignment
with its operator in the Hamiltonian. As before, the redundancy has
a linear region where it is proportional to $1-h/h_{m}$. This is
shown in Fig. \ref{fig:Rmismixed}(c).

\section{Conclusions\label{sec:Conclusions}}

We studied how information about a system of interest proliferates
throughout an environment under non-ideal initial conditions (namely,
hazy or misaligned initial environment states). When a system is undergoing
pure decoherence with a set of independent environment components,
we showed that, after decoherence has taken place, an environment
fragment's capacity to accept information about a system is given
by its ability to increase its entropy. Thus, increasing the overlap
of the environment with states that commute with the interaction Hamiltonian
(whether by misaligning it or by increasing its haziness) diminishes
its ability to increase its entropy and therefore decreases its capacity
to accept information about the system. Prior to the onset of good
decoherence, complementary information about the system (that is,
information about the superposition of pointer states of $\S$) is
transferred into the environment, where it is initially spread among
many fragments. After the onset of good decoherence, this complementary
information is encoded only globally in the environment (i.e., individual
fragments do not contain it) %
\footnote{This type of distribution of information may also be of interest in
other areas of research, such as representing environments in real-time
simulations \cite{Zwolak08-1}.%
}. Finally, we examined a model system of a symmetric qubit environment.
We found scaling relations that demonstrate a universal behavior of
the redundancy (i.e., behavior that is independent of the system's
initial state). Overall, our results show that although non-ideal
initial conditions diminish the environment's capacity to store information,
the environment still redundantly obtains information about the system
- demonstrating that Quantum Darwinism is robust and non-ideal environments
still communicate information redundantly.
\begin{acknowledgments}
We would like to thank Graeme Smith, Jon Yard, and Michael Zubelewicz.
This research is supported by the U.S. Department of Energy through
the LANL/LDRD Program.
\end{acknowledgments}
\appendix

\section{Qubit interacting with a symmetric environment\label{sec:AppQub}}

The total state of $\S\E$ evolves according to $\rho\left(t\right)=\U\left(t\right)\rho\left(0\right)\U^{\dg}\left(t\right)$,
where $\mathcal{U}\left(t\right)=\exp{\left(-\im\H t\right)}$ can
be written as \begin{equation}
\mathcal{U}\left(t\right)=\left\vert 0\right\rangle \left\langle 0\right\vert \otimes\left[\V\left(t\right)^{\otimes\Es}\right]+\left\vert 1\right\rangle \left\langle 1\right\vert \otimes\left[\V\left(-t\right)^{\otimes\Es}\right].\end{equation}
Here $\V\left(t\right)$ is the unitary matrix $\exp\left[-\im t\sigma^{z}/2\right].$
The evolution of $\S$ is given by \begin{eqnarray}
\rhoS & = & \left(\begin{array}{cc}
s_{00} & s_{01}\LE\\
s_{10}\LEc & s_{11}\end{array}\right),\label{eq:rhoS}\end{eqnarray}
with the total decoherence factor \begin{equation}
\LE=\prod_{k\in\E}\Lambda_{k}(t)\end{equation}
due to the environment $\E$. Each component of the environment contributes
a partial factor \begin{equation}
\Lambda_{k}\left(t\right)\equiv\mathrm{tr}\left[\bar{\rho}_{r}\left(t\right)\right]=\cos\left(t\right)-\im\sigma\sin\left(t\right)\label{eq:Lamk}\end{equation}
to the total decoherence. The state of $\S\F$ is \begin{equation}
\rhoSF=\left(\begin{array}{cc}
s_{00}\tilde{\rho}_{r}\left(t\right)^{\otimes\Fs} & s_{01}\bar{\rho}_{r}\left(t\right)^{\otimes\Fs}\LEF\\
s_{10}\bar{\rho}_{r}^{\dg}\left(t\right)^{\otimes\Fs}\LEFc & s_{11}\tilde{\rho}_{r}\left(-t\right)^{\otimes\Fs}\end{array}\right),\label{eq:rhoSF}\end{equation}
where $\tilde{\rho}_{r}\left(t\right)=\V\left(t\right)\rho_{r}\V^{\dg}\left(t\right)$
is a rotated density matrix on a single environment qubit and $\bar{\rho}_{r}\left(t\right)=\V\left(t\right)\rho_{r}\V\left(t\right)$
is an operator on a single environment qubit. 

The von Neumann entropy, $\ES$, can be calculated explicitly by diagonalizing
$\rhoS$ to obtain $H_{\S}(t)=H\left(\kE\right)$ with \begin{equation}
\kappa_{\mathcal{A}}(t)=\frac{1}{2}\left(1+\sqrt{\left(s_{11}-s_{00}\right)^{2}+4\left|s_{01}\right|^{2}\left|\Lambda_{\mathcal{A}}(t)\right|^{2}}\right).\label{eq:kapAApp}\end{equation}
The quantity $\kappa_{\mathcal{A}}(t)$ is one of the eigenvalues
of the state of $\S$ when it interacts only with the environment
components $k$ for which $k\in\mathcal{A}$. We can likewise readily
obtain the entropy of $\S\F$ by utilizing Eq. \eqref{eq:GDproof}:
this entropy is equivalent to the sum of the entropy of the system
decohered solely by $\E/\F$, i.e., the remainder of the environment,
which is given by $H\left(\kEF\right)$, and the entropy of the initial
state of the $\F$, $H_{\F}(0)=\Fs h$. Thus, the mutual information
becomes \begin{eqnarray}
I\left(\S:\F\right) & = & \left[H_{\F}\left(t\right)-H_{\F}\left(0\right)\right]\nonumber \\
 &  & +\left[H\left(\kE\right)-H\left(\kEF\right)\right].\end{eqnarray}
To finish the calculation of the mutual information, we need the remaining
term in Eq. \eqref{eq:MutInfo}: the entropy $\EF$. Generally, the
calculation of this entropy is difficult, as it requires diagonalizing
the reduced density matrix of $\F$, which in this case is \begin{equation}
\rhoF=s_{00}\tilde{\rho}_{r}\left(t\right)^{\otimes\Fs}+s_{11}\tilde{\rho}_{r}\left(-t\right)^{\otimes\Fs}.\label{eq:rhoF}\end{equation}
Due the symmetry of the problem, however, Eq. \eqref{eq:rhoF} can
be diagonalized efficiently numerically using the procedure outlined
in Appendix \ref{sec:RotTech}. Further, in the case of a pure initial
environment, one can compute $\EF$ analytically. In the following
subsections, we will examine several cases of how the mutual information
develops in time for different initial states.

\subsection{Pure or mixed $\S$ and pure $\E$}

When the environment is pure, the entropy of $\rhoF$ can be found
by purifying $\S$ using an ancillary system $\tilde{\S}$ and noting
that $\EF=\ESSEF$. Let $\lambda\equiv\left|s_{01}/\sqrt{s_{00}s_{11}}\right|$
parametrize the existing decoherence of $\S$. Purifying the initial
state of $\S$ gives \begin{equation}
\ket{\psi_{\S\tilde{S}}}=\alpha\ket{00}+\beta\ket{1\tilde{1}},\end{equation}
where $\left|\alpha\right|^{2}=s_{00}$, $\left|\beta\right|^{2}=s_{11}$,
and $\ket{\tilde{1}}=\lambda^{2}\ket 0+\sqrt{1-\lambda^{2}}\ket 1$
is a state of $\tilde{\S}$ that would give the existing decoherence
of $\S$. To calculate the entropy, $\EF=\ESSEF$, we can use Eq.
\eqref{eq:GDproof} with $\S$ replaced by $\S\tilde{\S}$ to show
that, in the presence of an initially pure $\E$ (and hence, $\E/\F$),
this entropy is equivalent to the entropy of $\S\tilde{\S}$ decohered
just by $\F$. The latter is \begin{eqnarray}
\rho_{\S\tilde{\S}}(t)= & s_{00}\proj{00}{00}+\sqrt{s_{00}s_{11}}\LF\proj{00}{1\tilde{1}}\nonumber \\
 & +\sqrt{s_{00}s_{11}}\LFc\proj{1\tilde{1}}{00}+s_{11}\proj{1\tilde{1}}{1\tilde{1}}.\end{eqnarray}
Since $\ket{00}$ and $\ket{1\tilde{1}}$ are orthogonal, the entropy
can be obtained from the eigenvalues of the matrix \begin{equation}
\left(\begin{array}{cc}
s_{00} & \sqrt{s_{00}s_{11}}\LF\\
\sqrt{s_{00}s_{11}}\LFc & s_{11}\end{array}\right),\end{equation}
which gives $H\left(\tilde{\kappa}_{\F}(t)\right)$, with \begin{equation}
\tilde{\kappa}_{\mathcal{A}}(t)=\frac{1}{2}\left(1+\sqrt{\left(s_{11}-s_{00}\right)^{2}+4s_{00}s_{11}\left|\Lambda_{\mathcal{A}}(t)\right|^{2}}\right).\end{equation}
Note that this result, $\EF=H\left(\tilde{\kappa}_{\F}(t)\right)$,
is indicating that the entropy of an initially pure $\F$ with time
is the same regardless of whether the system was initially pure or
mixed. Moreover, as we will see in just a moment, only the discord
changes when $\S$ is initially mixed. The mutual information is therefore
\begin{equation}
\MI=H\left(\tilde{\kappa}_{\F}(t)\right)+\left[H\left(\kE\right)-H\left(\kEF\right)\right],\label{eq:MIpureE}\end{equation}
where the last two terms in brackets give the quantum discord (and
the deviation from good decoherence) for an initially pure $\E$.

As a special case of the above, when $\S$ is pure $\tilde{\kappa}$
reduces to $\kappa$ in Eq. \eqref{eq:kapAApp} and the mutual information
is \begin{equation}
\MI=H\left(\kF\right)+\left[H\left(\kE\right)-H\left(\kEF\right)\right].\label{eq:MIpure}\end{equation}
This result can be found much more readily by using the equality $\EF=\ESEF$
for bipartite pure states. Then, employing Eq. \eqref{eq:ESF} for
$\ESEF$ and $\EEFo=0$ gives \begin{equation}
\EF=\EgF.\end{equation}
Thus we obtain $\EF=H\left(\kF\right)$. This shorter derivation for
initially pure $\S$ and $\E$ shows that the entropy of $\F$ is
simply the entropy of $\S$ when it is interacting solely with $\F$
\cite{Zurek07-1}.

\subsection{Pure or mixed $\S$ and hazy $\E$}

When the environment is hazy, the entropy of $\rhoF$ can not be found
by appealing to entropic properties of bipartite pure states, as was
done in the previous section. With our model, however, we can diagonalize
$\rhoF$ directly by taking advantage of the symmetry. By using the
Wigner D-matrices \cite{Cirac99-1,Dachsel06-1,Miyazaki07-1}, we can
rewrite $\rhoF$ into block diagonal form (see Appendix \ref{sec:RotTech}),
with a maximum block dimension equal to $2\Fs+1$. Thus, the complexity
for diagonalizing $\rhoF$ is reduced from exponential to polynomial
in $\Fs$.

In addition, we can also obtain an analytical result for the entropy
when $r_{00}=1/2$ and $\ts$. Under these two conditions, the reduced
density matrix of the environment becomes \begin{equation}
\rhoFp=s_{00}\left(\begin{array}{cc}
\frac{1}{2} & -\im r_{01}\\
\im r_{10} & \frac{1}{2}\end{array}\right)^{\otimes\Fs}+s_{11}\left(\begin{array}{cc}
\frac{1}{2} & \im r_{01}\\
-\im r_{10} & \frac{1}{2}\end{array}\right)^{\otimes\Fs}\label{eq:rhoFts}\end{equation}
At this time, both terms are diagonal in the same basis %
\footnote{For real $r_{01}$, this basis is given by the eigenstates of $\sigma^{y}$
for each of the qubits, see Fig. \ref{fig:Bloch}.%
}. Thus, the matrix can be diagonalized to yield \begin{equation}
\rhoFp=s_{00}\left(\begin{array}{cc}
\lambda_{+} & 0\\
0 & \lambda_{-}\end{array}\right)^{\otimes\Fs}+s_{11}\left(\begin{array}{cc}
\lambda_{-} & 0\\
0 & \lambda_{+}\end{array}\right)^{\otimes\Fs}\end{equation}
where $\lambda_{\pm}=1/2\pm\left|r_{01}\right|$. Its entropy is then
\begin{equation}
H_{\F}\left(\pi/2\right)=-\sum_{n=0}^{\Fs}\left(\begin{array}{c}
\Fs\\
n\end{array}\right)\lambda_{\F}\left(n\right)\log_{2}\left[\lambda_{\F}\left(n\right)\right],\label{eq:ExactSF}\end{equation}
where $\lambda_{\F}\left(n\right)=s_{00}\lambda_{-}^{n}\lambda_{+}^{\Fs-n}+s_{11}\lambda_{-}^{\Fs-n}\lambda_{+}^{n}$
are the degenerate eigenvalues of $\rhoFp$. The quantum discord at
this time is zero except when $\Fs=\Es$, thus the mutual information
is given exactly by Eq. \eqref{eq:GDMI} when $\Fs\neq\Es$. Since
$\EFo=\Fs\, h$, we obtain \begin{equation}
\MI=H_{\F}\left(\pi/2\right)-\Fs\, h.\label{eq:MIST}\end{equation}
In Appendix \ref{sec:Gaussian} we find an asymptotic approximation
to Eq. \eqref{eq:MIST}.

\begin{widetext}

\section{Diagonalizing $\rhoF$\label{sec:RotTech}}

A fragment $\F$ of the environment is described by the density matrix
(see Eq. \eqref{eq:rhoF}) \begin{equation}
\rhoF=s_{00}\tilde{\rho}_{r}\left(t\right)^{\otimes\Fs}+s_{11}\tilde{\rho}_{r}\left(-t\right)^{\otimes\Fs},\label{eq:rhoFApp}\end{equation}
where $\tilde{\rho}_{r}\left(t\right)=\V\left(t\right)\rho_{r}\V^{\dg}\left(t\right)$
is a rotated density matrix on a single environment qubit and the
initial density matrix is given by Eq. \eqref{eq:rhor}: \begin{equation}
\rho_{r}=\left(\begin{array}{cc}
r_{00} & r_{01}\\
r_{10} & r_{11}\end{array}\right).\label{eq:rhorApp}\end{equation}

To calculate the entropy of $\rhoF$, our strategy is to rewrite the
operators of the form $\tilde{\rho}_{r}\left(\pm t\right)^{\otimes\Fs}$
into direct sums of total spin states so that the density matrix becomes
block diagonal. Each block can then be diagonalized separately and
the computational cost of the computing the entropy is polynomial
in $\Fs$ rather than exponential. This process, which consists of
three steps, is illustrated in the schematic diagram shown in Fig.
\ref{fig:rotating}.

\begin{figure}[ht]
\begin{centering}
\includegraphics[width=10cm]{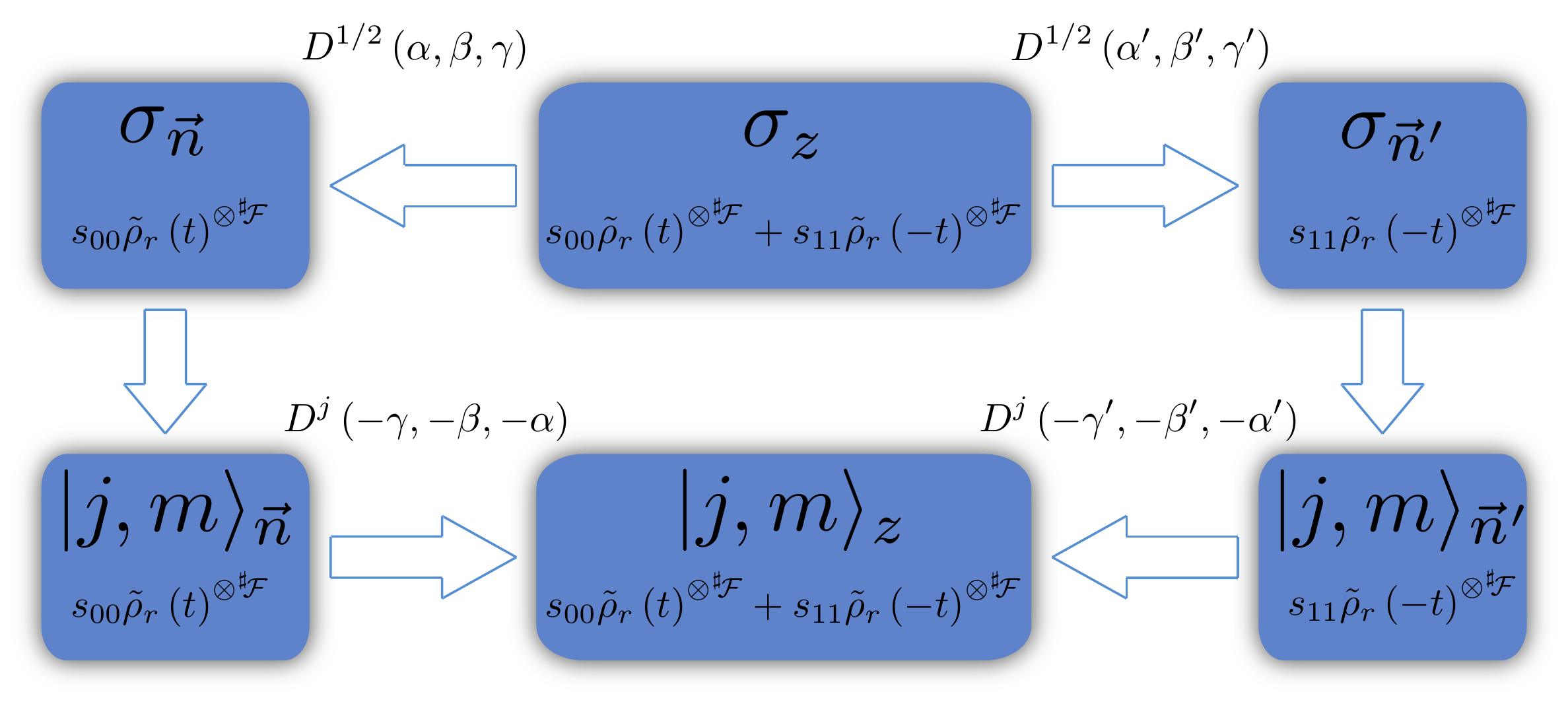}
\par\end{centering}

\caption{Schematic diagram of the rotating technique used to diagonalize the
density matrix $\rhoF$. The density matrix is split into two parts,
$s_{00}\tilde{\rho}_{r}\left(t\right)^{\otimes\Fs}$ and $+s_{11}\tilde{\rho}_{r}\left(-t\right)^{\otimes\Fs}$,
which are each separately rotated into the basis $\left\vert j,m\right\rangle _{z}$
by first going through the basis in which the state $\tilde{\rho}_{r}\left(t\right)$
is diagonal.\label{fig:rotating} }

\end{figure}

First, we make a unitary transformation to diagonalize $\tilde{\rho}_{r}\left(t\right)$
and $\tilde{\rho}_{r}\left(-t\right)$. This process can be alternatively
understood as a rotating of the density matrix $\tilde{\rho}_{r}\left(t\right)$
with a Wigner D-matrix \cite{Cirac99-1,Dachsel06-1,Miyazaki07-1},
$R(\alpha,\beta,\gamma)$, to change the representation from $\sigma_{z}$
to $\sigma_{\vec{n}}$, where $\vec{n}$ is the Bloch vector of the
spin. The second step is to rewrite $\tilde{\rho}_{r}\left(t\right)^{\otimes\Fs}$
into direct sums of total spin states by utilizing the Clebsch-Gordan
coefficients. After this step the representation is changed from $\sigma_{\vec{n}}$
to $\left\vert j,m\right\rangle _{\vec{n}}$. The third step is to
rotate from the representation $\left\vert j,m\right\rangle _{\vec{n}}$
to $\left\vert j,m\right\rangle _{z}$ by a inverse Wigner D matrix
$R(-\gamma,-\beta,-\alpha)\equiv R^{-1}(\alpha,\beta,\gamma)$. We
apply the rotating techniques separately to $\tilde{\rho}_{r}\left(t\right)^{\otimes\Fs}$
and $\tilde{\rho}_{r}\left(-t\right)^{\otimes\Fs}$, but finally bring
them both into the basis $\left\{ \left\vert j,m\right\rangle _{z}\right\} $
where the blocks are diagonalized.

The details of the procedure start with the rotation by the angles
$\alpha$, $\beta$, and $\gamma$: \begin{equation}
R(\alpha,\beta,\gamma)=e^{-i\alpha J_{z}}e^{-i\beta J_{y}}e^{-i\gamma J_{z}},\label{a3}\end{equation}
where $J_{x}$, $J_{y}$, and $J_{z}$ are the components of the angular
momentum (which for our spin system are just the Pauli matrices).
The Wigner D matrix is a square matrix of dimension $2j+1$ with general
element \begin{eqnarray}
D_{m,m^{\prime}}^{j} & = & \left\langle j,m^{\prime}\right\vert R(\alpha,\beta,\gamma)\left\vert j,m\right\rangle \label{a4}\\
 & = & e^{-im^{\prime}\alpha}d_{m^{\prime},m}^{j}(\beta)e^{-im\gamma},\end{eqnarray}
 where \begin{equation}
\begin{split}d_{m^{\prime},m}^{j}(\beta)= & \left\langle j,m^{\prime}\right\vert e^{-i\beta J_{y}}\left\vert j,m\right\rangle \\
= & [(j+m^{\prime})!(j-m^{\prime})!(j+m)!(j-m)!]^{\frac{1}{2}}\times\sum_{s=\mathrm{max}(0,m-m^{\prime})}^{\mathrm{min}(j+m,j-m^{\prime})}\frac{(-1)^{m^{\prime}-m+s}}{(j+m-s)!s!(m^{\prime}-m+s)!(j-m^{\prime}-s)!}\\
 & \times\left(\cos\left[\beta/2\right]\right)^{2j+m-m^{\prime}-2s}\left(\sin\left[\beta/2\right]\right)^{m^{\prime}-m+2s}.\label{a5}\end{split}
\end{equation}
The Euler angles $\alpha$, $\beta$, $\gamma$ in the rotation, Eq.
\eqref{a4}, are completely determined by the unitary matrix that
diagonalizes $\tilde{\rho}_{r}(t)$, $U\tilde{\rho}_{r}(t)U^{\dagger}=\mathrm{Diag}\left[\lambda_{+},\lambda_{-}\right]$,
which is \begin{equation}
U=\left[\begin{array}{cc}
\frac{-r_{01}e^{-it}}{\sqrt{|r_{01}|^{2}+(r_{00}-\lambda_{+})^{2}}}, & \frac{r_{00}-\lambda_{+}}{\sqrt{|r_{01}|^{2}+(r_{00}-\lambda_{+})^{2}}}\\
\frac{r_{11}-\lambda_{-}}{\sqrt{|r_{10}|^{2}+(r_{11}-\lambda_{-})^{2}}}, & \frac{-r_{10}e^{it}}{\sqrt{|r_{10}|^{2}+(r_{11}-\lambda_{-})^{2}}}\end{array}\right].\label{a6}\end{equation}
 This is equal to the Wigner D matrix \begin{equation}
D^{1/2}\left(\alpha,\beta,\gamma\right)=\left[\begin{array}{cc}
e^{-i(\alpha+\gamma)/2}\cos\left(\beta/2\right), & -e^{-i(\alpha-\gamma)/2}\sin\left(\beta/2\right)\\
e^{i(\alpha-\gamma)/2}\sin\left(\beta/2\right), & e^{i(\alpha+\gamma)/2}\cos\left(\beta/2\right).\end{array}\right]\label{a7}\end{equation}
 with the Euler angles\begin{equation}
\alpha=\gamma=t,\end{equation}
\begin{equation}
\sin\left(\beta/2\right)=-\frac{r_{00}-\lambda_{+}}{\sqrt{|r_{01}|^{2}+(r_{00}-\lambda_{+})^{2}}},\end{equation}
and \begin{equation}
\cos\left(\beta/2\right)=\frac{-r_{01}}{\sqrt{|r_{01}|^{2}+(r_{00}-\lambda_{+})^{2}}}.\end{equation}
The density matrix $\tilde{\rho}_{r}\left(t\right)^{\otimes\Fs}$
becomes \begin{equation}
\tilde{\rho}_{r}\left(t\right)^{\otimes\Fs}\rightarrow\mathrm{Diag}\left[\lambda_{+},\lambda_{-}\right]^{\otimes\Fs}\label{eq:DiagTen}\end{equation}
and similarly for $\tilde{\rho}_{r}\left(-t\right)^{\otimes\Fs}$.

Utilizing the Clebsch-Gordan coefficients, Eq. \eqref{eq:DiagTen}
can be rewritten as a direct sum of the total spin states, \begin{equation}
\mathrm{Diag}\left[\lambda_{+},\lambda_{-}\right]^{\otimes\Fs}\rightarrow\oplus_{j=0}^{\Fs/2}\left(M_{j}^{\oplus B_{j}}\right),\label{a9}\end{equation}
where \begin{equation}
M_{j}=\mathrm{Diag}[\lambda_{+}^{\frac{\Fs}{2}+j}\lambda_{-}^{\frac{\Fs}{2}-j},\lambda_{+}^{\frac{\Fs}{2}+j-1}\lambda_{-}^{\frac{\Fs}{2}-j+1},...,\lambda_{+}^{\frac{\Fs}{2}-j}\lambda_{-}^{\frac{\Fs}{2}+j}]\end{equation}
 and \begin{equation}
B_{j}=\left(\begin{array}{c}
\Fs\\
\Fs/2-j\end{array}\right)-\left(\begin{array}{c}
\Fs\\
\Fs/2-j-1\end{array}\right).\end{equation}
The basis of the density matrix $\tilde{\rho}_{r}\left(t\right)^{\otimes\Fs}$
is now $\left\{ \left\vert j,m\right\rangle _{\vec{n}}\right\} $,
and under the same procedure the density matrix $\tilde{\rho}_{r}\left(-t\right)^{\otimes\Fs}$
will be in the basis $\left\{ \left\vert j,m\right\rangle _{\vec{n}^{\p}}\right\} $
with the Bloch vector $\vec{n}^{\p}$. To get the full density matrix,
$\rhoF$, we need to transform them into the same basis $\left\{ \left\vert j,m\right\rangle _{z}\right\} $,
which can be done by rotating backwards using $D^{j}(-\gamma,-\beta,-\alpha)$
with the angles corresponding to the forward rotation $D^{1/2}(\alpha,\beta,\gamma)$:
\begin{equation}
\oplus_{j=0}^{\Fs/2}\left(M_{j}^{\oplus B_{j}}\right)\rightarrow\oplus_{j=0}^{\Fs/2}\left[e^{-i(-\gamma)J_{z}}e^{-i(-\beta)J_{y}}e^{-i(-\alpha)J_{z}}M_{j}e^{-i\alpha J_{z}}e^{-i\beta J_{y}}e^{-i\gamma J_{z}}\right]^{\oplus B_{j}}.\label{a10}\end{equation}
Now we can write $\rho_{\F}(t)=s_{00}\tilde{\rho}_{r}\left(t\right)^{\otimes\Fs}+s_{11}\tilde{\rho}_{r}\left(-t\right)^{\otimes\Fs}$
into a block diagonal form in the basis $\left\{ \left\vert j,m\right\rangle _{z}\right\} $,
which can be diagonalized efficiently to obtain the entropy of $\F$.

\section{Asymptotic approximation\label{sec:Gaussian}}

In this appendix, we approximate the expression in Eq. \eqref{eq:MIST}
for large $\Fs$. Our starting point is to rewrite Eq. \eqref{eq:MIST}
as \begin{eqnarray}
\MI & = & H_{\S}-s_{00}\sum_{n=0}^{\Fs}\left(\begin{array}{c}
\Fs\\
n\end{array}\right)\lambda_{-}^{n}\lambda_{+}^{\Fs-n}\log_{2}\left[1+\frac{s_{11}}{s_{00}}\left(\frac{\lambda_{-}}{\lambda_{+}}\right)^{\Fs-2n}\right]\nonumber \\
 &  & -s_{11}\sum_{n=0}^{\Fs}\left(\begin{array}{c}
\Fs\\
n\end{array}\right)\lambda_{-}^{\Fs-n}\lambda_{+}^{n}\log_{2}\left[1+\frac{s_{00}}{s_{11}}\left(\frac{\lambda_{+}}{\lambda_{-}}\right)^{\Fs-2n}\right]\\
 & \equiv & H_{\S}-\Delta\MI,\nonumber \end{eqnarray}
where we extracted out the plateau value of the mutual information,
$H_{\S}$, and also the initial entropy of $\F$, which cancelled
the second term in Eq. \eqref{eq:MIST}. The deviation of the mutual
information from its plateau value is defined as $\Delta\MI$, which
is the term we will approximate. For large $\Fs$, we can use the
de Moivre-Laplace theorem to replace the binomial coefficient: \begin{equation}
2^{\Fs}\left(\begin{array}{c}
\Fs\\
n\end{array}\right)\left(\frac{1}{2}\right)^{\Fs}\approx\frac{2^{\Fs}}{\sqrt{\pi\Fs/2}}e^{-\left(n-\Fs/2\right)^{2}/\left(\Fs/2\right)}.\end{equation}
Performing this replacement and rearranging some terms gives\begin{equation}
\Delta\MI\approx\frac{\left(2\sqrt{\lambda_{-}\lambda_{+}}\right)^{\Fs}}{\sqrt{\pi\Fs/2}}\sum_{n=0}^{\Fs}e^{-\left(n-\Fs/2\right)^{2}/\left(\Fs/2\right)}S\left(n\right),\end{equation}
where \begin{equation}
S\left(n\right)\equiv s_{00}\left(\frac{\lambda_{-}}{\lambda_{+}}\right)^{n-\Fs/2}\log_{2}\left[1+\frac{s_{11}}{s_{00}}\left(\frac{\lambda_{-}}{\lambda_{+}}\right)^{\Fs-2n}\right]+s_{11}\left(\frac{\lambda_{+}}{\lambda_{-}}\right)^{n-\Fs/2}\log_{2}\left[1+\frac{s_{00}}{s_{11}}\left(\frac{\lambda_{+}}{\lambda_{-}}\right)^{\Fs-2n}\right].\end{equation}
To see how the mutual information approaches the plateau for large
$\Fs$, we can make a further approximation by recognizing that the
function, $S\left(n\right)$, within the sum peaks at\begin{equation}
n=\frac{\Fs-\left(\ln\frac{s_{00}}{s_{11}}\right)/\left(\ln\frac{\lambda_{-}}{\lambda_{+}}\right)}{2}\end{equation}
and decays exponentially when away from this maximum at a length scale
independent of $\Fs$. When $\Fs$ is large enough, the Gaussian,
which has a width proportional to $\sqrt{\Fs}$, is approximately
constant where $S\left(n\right)$ is non-negligible. Thus, for large
$\Fs$, we approximate the Gaussian as a constant (with its value
set at its maximum) and obtain \begin{equation}
\Delta\MI\approx\frac{\left(2\sqrt{\lambda_{-}\lambda_{+}}\right)^{\Fs}}{\sqrt{\pi\Fs/2}}\sum_{n=0}^{\Fs}S\left(n\right).\end{equation}
This already gives the asymptotic behavior of the mutual information:
For large enough $\Fs$, the sum over $S\left(n\right)$ is independent
of $\Fs$ because of the exponential decay of $S\left(n\right)$ away
from its maximum. However, to remove the sum and obtain a compact
expression, we can approximate the sum over $S\left(n\right)$ by
an integral. When $\E$ is fairly hazy, $S\left(n\right)$ is smooth
as function of $n$ and this approximation is a good one (although,
it will have a finite relative error as $\Fs\to\infty$). Changing
the sum to an integral and extending the limits to infinity gives
the approximate deviation

\begin{equation}
\Delta\MI\approx\Delta I_{app}\left(\S:\F\right)=\frac{\left(2\sqrt{\lambda_{-}\lambda_{+}}\right)^{\Fs}}{\sqrt{\pi\Fs/2}}\int_{-\infty}^{\infty}dnS\left(n\right)=\frac{\left(2\sqrt{\lambda_{-}\lambda_{+}}\right)^{\Fs}}{\sqrt{\pi\Fs/2}}\frac{2\pi\sqrt{s_{00}s_{11}}}{\left(\ln2\right)\left(\ln\frac{\lambda_{+}}{\lambda_{-}}\right)},\end{equation}
which is the asymptotic approximation used within the paper. In Fig.
\ref{fig:Errors}(a) we plot this asymptotic approximation along with
the exact data for the deviation of the mutual information from its
plateau value. In Fig. \ref{fig:Errors}(b) we plot the relative error
\begin{equation}
\left|\frac{\Delta I_{app}\left(\S:\F\right)-\Delta\MI}{\Delta\MI}\right|.\end{equation}
As can been seen from the figures, the asymptotic approximation correctly
describes the decay of the mutual information to its plateau value.
\begin{figure}
\begin{centering}
\includegraphics[width=12cm]{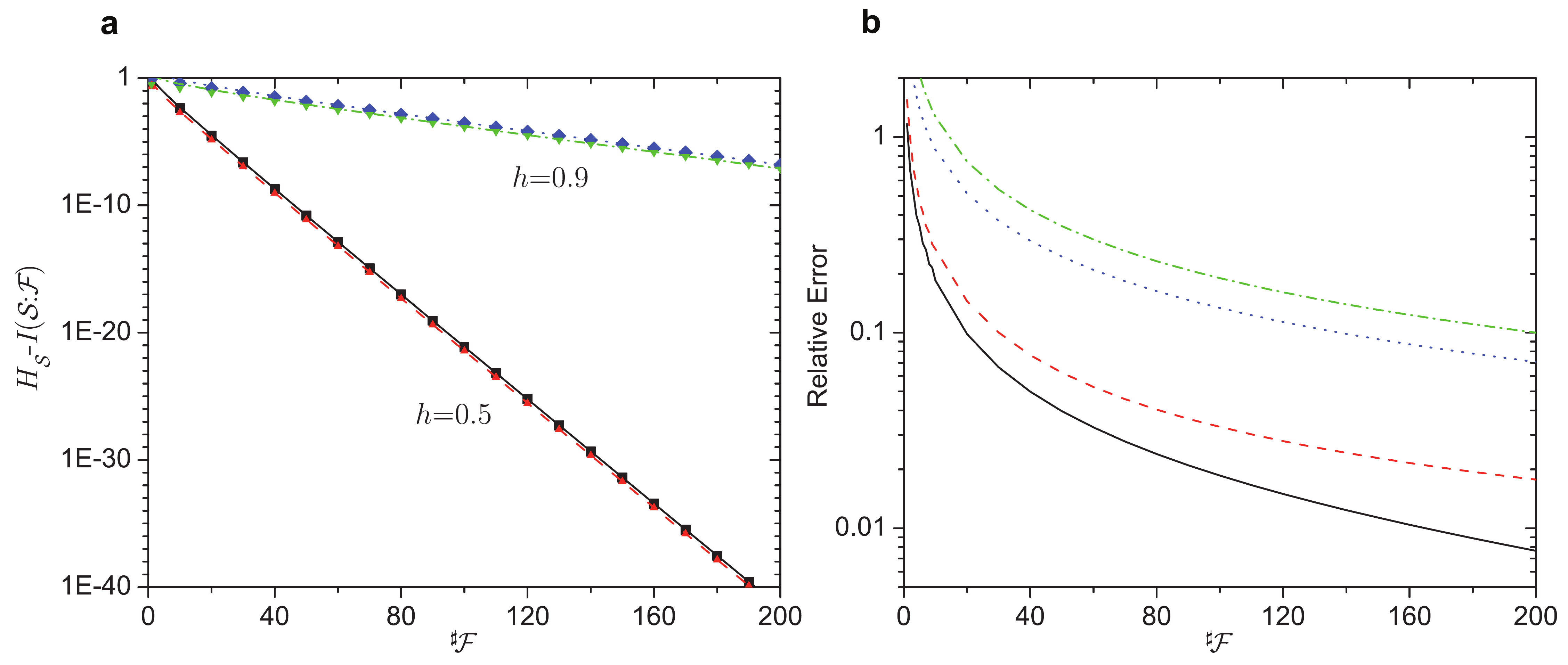}
\par\end{centering}

\caption{(a) Deviation of the mutual information from its plateau value, $\Delta\MI=H_{\S}-\MI$,
versus the fragment size $\Fs$. The exact deviation is plotted for
$h=0.5$, $s_{00}=1/2$ (black squares); $h=0.5$, $s_{00}=1/16$
(red triangles); $h=0.9$, $s_{00}=1/2$ (blue diamonds); and $h=0.9$,
$s_{00}=1/16$ (green inverse triangles), with the approximate data
plotted as a line of the same color as its corresponding exact data.
For all but the smallest $\Fs$, the approximation gives the correct
decay of the mutual information to its plateau value. Further, changing
the value of $H_{\S}$ (by shifting $s_{00}$) does not change the
decay behavior to the plateau. (b) Relative error of the asymptotic
approximation versus $\Fs$. The errors are for $h=0.5$, $s_{00}=1/2$
(black line); $h=0.5$, $s_{00}=1/16$ (red dashed line); $h=0.9$,
$s_{00}=1/2$ (blue dotted line); and $h=0.9$, $s_{00}=1/16$ (green
dash-dotted line). The errors decay initially as the approximation
of the binomial coefficient by a constant becomes better, but the
approximation will contain a finite relative error as $\Fs\to\infty$
due to the approximation of the sum by an integral.\label{fig:Errors}}

\end{figure}

\end{widetext}

\end{document}